\documentclass[usenatbib,useAMS]{mn2e}
\usepackage{epsfig,psfig,graphicx,float,color}
\usepackage{subfigure}
\usepackage{mybibdefs}
\usepackage{multirow}
\usepackage{widetext}
\usepackage{url}
\usepackage{bm}
\usepackage{float}
\usepackage{color}
\usepackage{amsmath}
\usepackage{mathtools}
\usepackage{listings}% http://ctan.org/pkg/listings
\usepackage{fancybox}


\newcommand{\rc}{\textcolor{black}}

\usepackage{graphicx}
\usepackage{amssymb}
\usepackage{epstopdf}
\title[Accurate photometric redshift PDF estimation]{ Accurate photometric redshift probability density estimation - method comparison and application}
\author[Rau et al.]{Markus Michael Rau$^{1,2}$, Stella Seitz$^{1,2}$, Fabrice Brimioulle$^3$, Eibe Frank$^4$,\newauthor Oliver Friedrich$^{1,2}$, Daniel Gruen$^{1,2}$, Ben Hoyle$^{1,5}$ \\\\\\\\
$^1$Ludwig-Maximilians-Universit\"at M\"unchen, Universit\"ats-Sternwarte, Scheinerstr. 1, D-81679 Munich, Germany\\ 
$^2$Max-Planck-Institut f\"ur extraterrestrische Physik, Giessenbachstrasse 1, 85748 Garching, Germany\\
$^3$Observat\'{o}rio Nacional, Rua Gal. Jos\'{e} Cristino, 20921-400, Rio de Janeiro, Brazil\\
$^4$Department of Computer Science, University of Waikato, Hamilton, New Zealand\\
$^5$Excellence Cluster Universe, Bolzmannstr. 2, D-85748 Garching, Germany\\
{\tt E-mail: mmrau@usm.lmu.de}
}

\begin{document}
\date{Accepted ----. Received ----; in original form ----.}
\pagerange{\pageref{firstpage}--\pageref{lastpage}} \pubyear{2014}
\maketitle
\label{firstpage}
\begin{abstract}
We introduce an ordinal classification algorithm for photometric redshift estimation, which significantly improves the reconstruction of photometric redshift probability density functions (PDFs) for individual galaxies and galaxy samples. As a use case we apply our method to CFHTLS galaxies. The ordinal classification algorithm treats distinct redshift bins as ordered values, which improves the quality of photometric redshift PDFs, compared with non-ordinal classification architectures. We also propose a new single value point estimate of the galaxy redshift, that can be used to estimate the full redshift PDF of a galaxy sample. This method is competitive in terms of accuracy with contemporary algorithms, which stack the full redshift PDFs of all galaxies in the sample, but requires orders of magnitudes less storage space.

The methods described in this paper greatly improve the log-likelihood of individual object redshift PDFs, when compared with a popular Neural Network code (ANNz). In our use case, this improvement reaches 50\% for high redshift objects ($z \geq 0.75$). 

We show that using these more accurate photometric redshift PDFs will lead to a reduction in the systematic biases by up to a factor of four, when compared with less accurate PDFs obtained from commonly used methods. The cosmological analyses we examine and find improvement upon are the following: gravitational lensing cluster mass estimates, modelling of angular correlation functions, and modelling of cosmic shear correlation functions.
\end{abstract}

\section{INTRODUCTION}
\label{introduction}
The determination of distance, or redshift, estimates to galaxies is a vital requirement before using large scale photometric galaxy surveys for many cosmological analyses. Large scale surveys, such as the SDSS \citep{SDSS_reference},  PanSTARRS \citep{PanSTARRS}, DES \citep{DES_ref} and LSST \citep{LSST_reference} rely on a combination of photometric and more accurate spectroscopic redshifts when providing distance estimates to photometrically identified galaxies.

Photometric redshifts are used throughout astrophysics and cosmology, for example in large scale structure analyses \citep{annz2, annz1}, in galaxy cluster weak lensing analyses \citep{GruenBrim2013}, and in galaxy-galaxy lensing analyses \rc{\citep{Brimioulle2013}}. Photometric redshifts are obtained using either machine learning methods or template fitting techniques \citep[see e.g.,][]{Benitez2000, Csabai2000, Bender2001, Ilbert2006, ZebraRef, Greisel2013}. Machine learning techniques range from early works employing artificial Neural Networks \citep{nnearlyreference, collister} as photometric point predictors, to recent developments that estimate the full photometric redshift PDF of the galaxy \citep{Lima2008, cunha, tpz, bonnet}. For detailed reviews and comparisons of different photometric redshift techniques we refer the reader to \cite{dessanchez, hildebrand_comparison, Dahlen_comparison}. This work focuses on machine learning methods for photometric redshift PDF estimation for samples of galaxies (hereafter sample PDF) as well as individual galaxies (hereafter individual PDFs). We apply the results to a range of analyses in weak gravitational lensing, cosmic shear and large scale structure. 

In general, machine learning algorithms learn a mapping between the photometry of an object and the spectroscopic redshift. To train the machine learning models to learn this mapping, one typically identifies spectrophotometric data that overlaps with the photometric feature space of the final data sample for which one would like to estimate redshifts. However, recent work shows that machine learning can also be performed with spectroscopic reference data that is brighter than the photometric sample \citep{Hoyle2015}. Many photometric surveys include a dedicated spectroscopic follow up program, which allows such a machine learning system to be built, e.g., SDSS-I/II \citep{York2000}, 2dF \citep{2dF_survey}, VVDS \citep{VVDS}, WiggleZ \citep{WiggleZ_ref}. 

The mapping obtained with machine learning is only approximate: the redshift of an object cannot be exactly determined by its corresponding photometry. Moreover, most machine learning methods produce a point estimate, which reduces the individual PDF to one number. The point estimate only predicts the most likely value of the redshift, irrespective of the quality of the photometry, and the shape of the distribution. In order to enter the era of precision cosmology, one must be able to incorporate the uncertainty in the redshift estimate into the cosmological analysis. This means that the use of single point redshift predictions is no longer sufficient. To achieve precision cosmology, we are required to incorporate the full redshift uncertainty using the individual PDFs. 

We can obtain a sample PDF by stacking the individual PDFs. This distribution describes the probability that a randomly sampled galaxy has a certain redshift. The accurate estimation of the redshift distribution of the full sample is important for many cosmological analyses, e.g, in large scale structure, weak gravitational lensing, and cosmic shear. 

However, effectively estimating and storing the photometric redshift PDF instead of the point estimate, for each object in a large astronomical dataset, is a challenging task. This process requires efficient and accurate photometric estimation algorithms, and scalable data storage solutions. These algorithms must be benchmarked using carefully constructed performance metrics to be useful for the next generation large scale structure photometric surveys \citep[e.g.,][]{EuclidReport}.

We discuss such metrics to quantify performance of photometric redshift PDF estimation in \S \ref{concept}.
We describe the Ordinal Class PDF (OCP) algorithm in \S \ref{subsec:ocp}, which improves the estimation accuracy over commonly used non-ordinal classification architectures. We continue in \S \ref{subsec:fit_gmm} by showing how the OCP method can become more storage efficient, by combining it with the Gaussian mixture model. This enables the storage of the PDFs of individual galaxies even within massive datasets without significant demands on disc space.

Many applications in cosmology require an estimation of the sample PDF. We propose a single point estimator for this quantity in \S \ref{subsec:hwe}, and show how this single floating point number can be computed very efficiently, and achieves good performance when compared with algorithms that stack individual PDFs. The performance of the proposed techniques is demonstrated and analysed in a method comparison in \S \ref{subsub:comp_annz} and \S \ref{subsec:smnlp} using a spectrophotometric dataset (\S \ref{dataset}) obtained from the public CFHTLS WIDE survey.

Finally, we demonstrate in \S \ref{subsec:application} that the methods introduced in this work improve the precision of gravitational lensing cluster mass estimates, measurements of angular correlation functions, and analyses of cosmic shear correlation functions, when compared with results obtained using a common Neural Network code. We conclude and summarize in \S \ref{conclusions}.

\section{FUNDAMENTAL CONCEPTS}
\label{concept}
The following section gives a brief review of important statistical concepts needed in this work. We start with a short introduction to density estimation, introduce metrics to quantify the performance of density estimators and finally describe a scheme to assess the performance of a machine learning model. 
\subsection{Kernel Density Estimation}
\label{subsec:kde}
The goal of kernel density estimation is to find a good estimator\footnote{In the following we will mark the estimator for a quantity with a hat.} $\hat{p}(\mathbf{x})$ for the probability density function $p(\mathbf{x})$ of a random variable $\mathbf{X}$ using N samples $\mathbf{x}_i$.  
Consider a small region $\mathcal{R}$ centred on a point $\mathbf{x}$. We can then assume that $p(\mathbf{x})$ is approximately constant across $\mathcal{R}$.
Based on this assumption we can estimate the density at point $\mathbf{x}$ as 
\begin{equation}
 \hat{p}(\mathbf{x}) = \frac{k}{N V_{\mathcal{R}}} \, .
 \label{eq:densestim}
\end{equation}
The number of objects\footnote{Fixing the number of points \textit{k} that fall into $\mathcal{R}$ and estimating the volume $V_{\mathcal{R}}$ leads to the \textit{k} nearest neighbour density estimation technique \citep[see e.g.][]{Scott1991}.} \textit{k} in Eq. \ref{eq:densestim} can be estimated by considering a $D$ dimensional hyper cube with volume
\begin{equation}
V_{\mathcal{R}} = h^D 
\end{equation}
centred on the point $\mathbf{x}$ with side length $h$. Using Eq. \ref{eq:densestim}, we obtain $k$ as
\begin{equation}
 k = \sum_{i = 1}^{N} K\left(\frac{\mathbf{x} - \mathbf{x}_i}{h} \right) \, ,
\end{equation} where 
\begin{equation}
 K(\mathbf{d}) = \begin{cases}
                  1,   &|d_i| \leq 1/2, \ \ 1 \leq i \leq D \\
                  0,   & \text{otherwise}
                 \end{cases} 
\label{eq:naive_kern}
\end{equation}
is an example of a kernel function. Note that this kernel has discontinuities at the boundaries. The bandwidth \textit{h} determines how much the kernel density estimate interpolates (or smoothes) between the given data points. A bandwidth that is too large oversmooths important structures in the density whereas one that it too small leads to a noisy density estimate.
The density estimate $\hat{p}(\mathbf{x})$ can then be written as 
\begin{equation}
 \hat{p}(\mathbf{x}) = \frac{1}{N} \sum_{i = 1}^{N} \frac{1}{h^{D}} 
K\left(\frac{\mathbf{x} - \mathbf{x}_i}{h} \right) =: \frac{1}{N} \sum_{i = 1}^{N} 
\tilde{K}\left(\mathbf{x}, \mathbf{x}_i, h \right) \, .
 \label{eq:kern_dens_def}
\end{equation}
Instead of using Eq. \ref{eq:naive_kern}, which has discontinuities 
at the boundaries, we can alternatively use smooth and symmetric functions, for example, a Gaussian. 

The estimation of photometric redshift PDFs for individual objects (individual PDFs) is an application of conditional probability density function estimation, since the individual PDF $p(z | \mathbf{f})$ is conditional on the objects photometry $\mathbf{f}$. The estimation of conditional probability density functions can be formulated in close analogy to Eq. \ref{eq:kern_dens_def}. We can estimate the individual PDF $p(z | \mathbf{f})$ as a weighted kernel density estimate in redshift space of the form
\begin{equation}
	\hat{p}(z | \mathbf{f}) = \sum_{i = 1}^{N_{\rm tr}} w_i(\mathbf{f}) K(z, z_{i}^{\rm spec}, h) \, ,
	\label{eq:cond_prob}
\end{equation}
using a dataset, the so called training set, containing $N_{\rm tr}$ objects. $K(z, z_{i}^{\rm spec}, h)$ denotes a kernel function with bandwidth \textit{h} centred on the spectroscopic redshift values $z_{i}^{\rm spec}$. The weights $w_i(\mathbf{f})$ sum to unity and depend on the photometry $\mathbf{f}$ of the object. 

The conditional cumulative distribution function $F(z | \mathbf{f})$ defined as 
\begin{equation}
	F(z | \mathbf{f}) = \int_{- \infty}^{z} p(z' | \mathbf{f}) dz'
	\label{eq:cond_ecdf}
\end{equation}
can be estimated \citep{Meinshausen2006} as 
\begin{equation}
	\hat{F}(z | \mathbf{f}) = \sum_{i = 1}^{N_{\rm tr}} w_i(\mathbf{f}) I(z_{i}^{\rm spec} \leq z) \, .
	\label{eq:estim_cond_ecdf}
\end{equation}
$I(z_{i}^{\rm spec} \leq z)$ equates to unity if $z_{i}^{\rm spec} \leq z$ and to zero otherwise.

The redshift distribution $\hat{p}(z)$ of a sample (sample PDF) containing \textit{N} objects can be estimated by stacking the individual PDFs
\begin{equation}
	\hat{p}(z) = \sum_{i = 1}^{N} w_{{\rm stack}, i} \, \hat{p}(z | \mathbf{f}_i) \, .
	\label{eq:stack_single_pdf}
\end{equation}

The normed weights $w_{{\rm stack}, i}$ can be set to $1/N$ or chosen to give more weight to certain sub populations. 
For example, we can favour certain redshift intervals $z \in [a, b]$ by defining weights as
\begin{equation}
	w_{{\rm stack}} = \int_{a}^{b} p(z | \mathbf{f}) dz =  \hat{F}(b | \mathbf{f}) - \hat{F}(a | \mathbf{f}) \, ,
	\label{eq:stack_weights}
\end{equation}
and we show an example of such a weighting in \S \ref{subsec:smnlp}.
The above weights are normalized afterwards to sum to unity.
\subsection{The Gaussian Mixture Model}
In this paper, we consider kernel density estimators and Gaussian mixture models for density estimation. A Gaussian mixture model \citep[see, for example,][]{Bishop2006} for the probability density function $p(x)$ of a random variable $X$ is a linear combination of $K$ normal densities defined as
\begin{equation}
	p(x) = \sum_{i = 1}^{K} \alpha_{i} \mathcal{N}\left(x, \mu_i, \sigma_i\right)
	\label{eq:gmm}
\end{equation}
where $\alpha_{i}$ is the amplitude, $\mu_i$ is the mean, and $\sigma_i$ is the standard deviation of the mixture component \textit{i}. 

We define the weight proportion $\gamma_{k}(x)$ of component \textit{k} as
\begin{equation}
\gamma_k(x) = \frac{\alpha_k \mathcal{N}\left(x, \mu_k, \sigma_k\right)}{\sum_{j = 1}^{K} \alpha_{j} \mathcal{N}\left(x, \mu_j, \sigma_j\right)} \, ,
	\label{eq:response}
\end{equation}
where $\gamma_{k}(x)$ determines how much a certain component of the Gaussian mixture model contributes to the total density at point $x$.
\subsection{Evaluation Metrics}
\label{subsec:eval_metrics}
Consider an estimate $\hat{p}(\mathbf{x})$ of the true probability density function $p(\mathbf{x})$ describing the distribution of the random variable $\mathbf{X}$. We can measure the quality of the estimate $\hat{p}(\mathbf{x})$ by its distance $D(\hat{p}(\mathbf{x}) || p(\mathbf{x}))$ to the true distribution $p(\mathbf{x})$, which is generally unknown.
The Kullback-Leibler divergence between the true density $p(\mathbf{x})$ and the estimate $\hat{p}(\mathbf{x})$ is defined using the natural logarithm as,
\begin{equation}
	D(p || \hat{p}) = \int_{-\infty}^{\infty} p(\mathbf{x}) \log{\left(\frac{p(\mathbf{x})}{\hat{p}(\mathbf{x})} \right)} \, d\mathbf{x} \, .
\end{equation}
A good estimate $\hat{p}$ for $p$ should minimize $D(p || \hat{p})$. Rewriting the logarithm we obtain
\begin{equation}
	D(p || \hat{p}) = \int_{-\infty}^{\infty} p(\mathbf{x}) \log{\left(p(\mathbf{x})\right)} \; \mathrm{d}\mathbf{x} - \int_{- \infty}^{\infty} p(\mathbf{x}) \log{\left(\hat{p}(\mathbf{x})\right)} \; \mathrm{d}\mathbf{x} \, ,
	\label{eq:kullback_expand}
\end{equation}
and we note that the first term is a constant that does not depend on the model parameters, for example bandwidth, kernel or shape of kernel function. Thus, the second term in Eq. \ref{eq:kullback_expand} can be used as a relative measure of the accuracy of $\hat{p}(\mathbf{x})$. If we use the sample mean to estimate the expectation with respect to $p(\mathbf{x})$, we obtain the mean negative log likelihood loss, hereafter MNLL, \citep{Habbema1974, Duin1976}
\begin{equation}
	MNLL = - \frac{1}{N} \sum_{i = 1}^{N} \log{\left(\hat{p}(\mathbf{x}_i) + \epsilon\right)} \, ,
	\label{eq:mnlp_orig}
\end{equation}
where we set $\epsilon = 10^{-6}$ to avoid floating point underflow.
The Kullback-Leibler Divergence is a distance and thus non negative and it is smallest if the MNLL is smallest.

A suitable loss function for individual PDFs can be defined analogously \citep[see e.g.,][]{Takeuchi2007, Frank2009, Sugiyama2010}. 
We estimate $p(z | \mathbf{f}_i)$ for each of the \textit{N} objects in the sample, in order to establish performance using a sample of objects for which spectroscopic redshift values have been observed. We then evaluate $\hat{p}(z | \mathbf{f}_i)$ at the object's observed spectroscopic redshift $\hat{p}(z = z_{{\rm spec}, i} | \mathbf{f}_i)$.
In the rest of the paper, the abbreviation MNLL refers to the mean negative log-likelihood loss evaluated for individual PDFs.
\subsection{Model Training}
\label{eq:testingframe}
We randomly sample three non-overlapping datasets without replacement from the available data: the training set, the validation set and the test set.
The model is trained on the training set and the model parameters are chosen by testing the performance of the trained model with different parameter settings on the validation set. 

The validation set is used during model tuning and therefore it does not provide a good estimate of the performance on unseen data. We measure this generalization performance on a test set that is not used during training and tuning. 

To evaluate the machine learning algorithms, we construct a training set containing 9000 objects, a validation set containing 3000 objects and a test set containing 22072 objects. After the validation set has been used to determine the best combination of model parameters, we merge the training set and the validation set and train the respective model again with this best setup. In this way, we make optimal use of the available data to build the final model. All results described in \S \ref{results} are obtained on the test set, which, we reiterate, was not used in all prior steps of model training and tuning.
\begin{table*} 
\centering
\begin{tabular}{|l||l|l|l|l|l}
    & QRF/HWE & OCP & NOCP &  OCP GMM  \\\hline\hline
nodesize & 3,\textbf{5},7,10  & 1,2,3,\textbf{5},7,9  & \textbf{1},2,3,5  & \textbf{1},2,3,5,7,9       \\
mtry & 1,2,3,4,\textbf{5}     & 1,2,3,\textbf{4},5  & 1,2,3,4,\textbf{5} & 1,2,3,4,\textbf{5}  \\
BW mod & 0.5,0.6,...,\textbf{1.8},...,3.0 & 0.5,0.6,...,\textbf{2.5},...,3.0 & 0.5,0.6,...,\textbf{2.0},...,3.0 & -  \\
Gauss Comp. & - & - & - & \textbf{1},2,3    \\ 
\end{tabular}
\caption[Tuning parameters]{Model parameters of the Quantile Regression Forest (QRF), the classification based PDF estimation algorithms (OCP/NOCP) and the OCP algorithm used with the Gaussian mixture model OCP GMM. `nodesize' and `mtry' are model parameters of the Random Forest described in \S \ref{algorithms}. `BW mod' is the bandwidth modification factor employed in the Scott's rule (Eq. \ref{eq:bw_select_scott}) and Gauss Comp. denotes the maximum number of components allowed in the Gaussian mixture model. The best parameter configuration for each algorithm picked on the validation set during model tuning (\S \ref{eq:testingframe}) is marked in bold type.}
\label{tab:model_select_pdf_comparison}
\end{table*}

In this work we choose to use the aperture magnitudes of the CFHTLS WIDE five band photometry as input attributes. Other photometric features may be used, for example see \citet{Hoyle2014} for a feature importance analysis.

\section{ALGORITHMS}
\label{algorithms}
We have introduced the estimator for the photometric redshift PDF of individual objects (individual PDF) in Eq. \ref{eq:cond_prob} as a weighted kernel density estimate that depends on the weights $w(\mathbf{f})$. The following section discusses two algorithms that can be used to estimate these weights.
\subsection{Quantile Regression Forest (QRF)}
\label{subsec:qrf}
The Quantile Regression Forest\footnote{The method was originally developed to estimate conditional quantiles hence the name Quantile Regression Forest.} \citep{Meinshausen2006} is a generalization of the Random Forest \citep{Breiman2001} that can be used to reconstruct individual PDFs, an algorithm known as TPZreg \citep{tpz} in astrophysics. 

A regression/classification tree partitions the input space and returns the mean/majority vote of the response values (i.e., the redshift values) of the training set objects in each partition as the final prediction for new objects falling into that partition. The tree partitions the input data such that the training set objects in each partition are most similar with respect to their response values. In regression, we measure similarity using the sum of squares loss function $SSE$, defined as
\begin{equation}
	SSE = \sum_{\tau = 1}^{l} \sum_{\mathbf{f}_i \in \mathcal{R}_{\tau}} \left(z_{{\rm spec}, i} - \left\langle z_{{\rm spec}, \tau} \right\rangle \right)^2 \, .
	\label{eq:sse_tree}
\end{equation}
The sum runs over all $l$ leaf nodes of the tree $1 \leq \tau \leq l$, which each represents a certain partition $\mathcal{R}_{\tau}$ in input space, and over all objects in the training set $(\mathbf{f}_i, z_{{\rm spec}, i})$ with attribute values $\mathbf{f}_i$ that fall into $\mathcal{R}_{\tau}$. The term $\left\langle z_{{\rm spec}, \tau} \right\rangle$ denotes the mean spectroscopic redshift of all training set objects that fall into $\mathcal{R}_{\tau}$.

The binary tree is recursively grown by choosing a splitting attribute and split point for each region using brute-force search such that the SSE is minimized. 

The Random Forest algorithm combines several trees by bootstrap aggregation which is described as follows. New training sets are drawn from the original training set with replacement, which is also known as bootstrapping. We train a tree model on each of these bootstrapped training sets, to obtain an ensemble of trees. Combining the estimates from all trees in the ensemble reduces variance. In addition, the Random Forest algorithm makes the resulting models even more diverse by modifying the way each tree is grown. Before each split selection, the routine randomly selects a certain number of attributes, as specified by the `mtry' parameter, on which the algorithm can perform the split. 

The complexity of the tree model is governed by the size of the leaves of the tree. We stop the recursive tree building process when a specified minimum number of objects in each leaf, denoted as `nodesize' is reached. If the nodesize is small, very complex trees are grown and the tree might overadapt to the training set. This is an example of overfitting. The prediction from the Random Forest is the mean, in regression, or the majority vote, in classification, of the predictions from the ensemble of trees. 

A single tree in the Random Forest splits the space spanned by the input attributes derived from the photometry of the objects into partitions which are represented by the tree leaves. Each leaf defined in this manner is associated with the mean spectroscopic redshift value of the training set objects in this leaf. The tree therefore approximates the underlying smooth function by a step function. If a new object is queried, it will be placed in a leaf containing objects with similar photometry. Following the formulation by \citet{Meinshausen2006}, we can write the photometric redshift prediction 
\begin{equation}
z_{\rm phot}(\mathbf{f}) = \sum_{i = 1}^{N_{\rm tr}} w_{i}(\mathbf{f}) z_{{\rm spec}, i}
\label{eq:cond_mean_pred}
\end{equation}
as a weighted sum over the spectroscopic redshift values $z_{{\rm spec}, i}$ of the $N_{\rm tr}$ training set objects. In order to distinguish the different trees in the ensemble, which are characterized by different split selections, we introduce a parameter $\mathbf{\theta}$, which characterizes each tree. All training set objects with photometry $\mathbf{f}_{i}^{\rm tr}$ that are located in the same region $\mathcal{R}_{l(\mathbf{f}, \theta)}$ (defined by the leaf $l(\mathbf{f}, \theta)$) as the newly queried object with photometry $\mathbf{f}$, get a constant weight, and all other training set objects get zero weight. This can be written as
\begin{equation}
	 w_{i}(\mathbf{f}, \theta) = \frac{I\left(\mathbf{f}_{i}^{\rm tr} \in \mathcal{R}_{l(\mathbf{f}, \theta)}\right)}{\sum_{j = 1}^{N_{\rm tr}} I\left(\mathbf{f}_{j}^{\rm tr} \in \mathcal{R}_{l(\mathbf{f}, \theta)}\right)} \, ,
\end{equation}
where the weights are normalized such that they sum to unity.  

The same concept holds for the Random Forest prediction, in which the weights associated with each training set object are averaged over $k$ trees, each grown on different bootstrapped datasets, and therefore each described by a different parameter $\mathbf{\theta}_{b}$:
\begin{equation}
	w_{i}(\mathbf{f}) = \frac{1}{k} \sum_{b = 1}^{k} w_{i}(\mathbf{f}, \theta_{b}) \, .
	\label{eq:QRF_weights}
\end{equation}
The weights can be used to estimate the individual PDF and corresponding statistics like the conditional mean, the conditional cumulative distribution function or the conditional standard deviation defined as
\begin{equation}
	\hat{\sigma}^2(z | \mathbf{f}) = \sum_{i = 1}^{N_{\rm tr}} w(\mathbf{f}_i) \left(z_{{\rm spec}, i} - z_{\rm phot}(\mathbf{f}_i)\right)^2 \, .
	\label{eq:standdev}
\end{equation}
The following section introduces an alternative way of estimating the weights in Eq. \ref{eq:cond_prob}, using a classification scheme.
\subsection{Ordinal Class PDF (OCP) estimation}
\label{subsec:ocp}
The basic idea of classification-based PDF estimation is to bin the spectroscopic data by redshift and use a classification algorithm that outputs probabilities for bin membership to reconstruct the PDF. Bin membership is viewed as an ordinal variable.
Ordinal scale variables, in contrast to nominal ones, exhibit an intrinsic order. If the classes in a classification problem are ordinal, we can use this information to improve the classification \citep{Frank2001}. 

Current classification-based PDF estimation methods in the astrophysics literature \citep[e.g.,][]{bonnet, tpz} treat redshift bins as nominal classes. In the following, we will refer to the latter as the non-ordinal class PDF (NOCP) algorithm.
\begin{figure}
   \centering
 \includegraphics[scale=0.33]{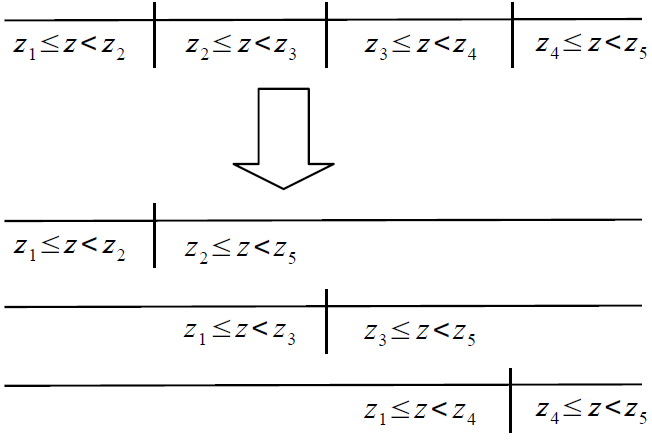}
   \caption{ \label{ordinal_class_trafo} An illustrative example of a nominal classification problem with four redshift bins. These bins can be reformulated into three binary classification problems by merging neighbouring bins. The class probabilities from the binary classification problems can be recombined to incorporate the ordering between the redshift bins (see text) into the final classification.  } 
\end{figure}
The ordinal class PDF (OCP) algorithm trains a separate classifier that estimates the probability $p(z \geq z_i)$ that a new object has redshift $z$ above a certain threshold $z_i$ given by the edge of the respective redshift bin. This scheme is illustrated in Fig. \ref{ordinal_class_trafo}.
The probability that the redshift of an object resides in the original bins is then calculated from these separate classification models as \citep{Frank2001} 
\begin{lstlisting}[numbers = left,xleftmargin=0.075\textwidth]
$p(z \in [z_{1}, z_{2}[) = 1 - p(z \geq z_2)$
$p(z \in [z_{i - 1}, z_{i}[) = p(z \geq z_{i - 1}) - p(z \geq z_i)$
$p(z \in [z_{k - 1}, z_{k}[) = p(z \geq z_{k - 1})$ $,$   $1 < i < k \, .$
\end{lstlisting}
The reconstruction of the class probabilities $p(z_i)$ has the idealistic assumption that each of the classifiers used to estimate the probability $p(z \geq z_i)$ outputs perfect probabilities. In practice, this will not be the case and the recovered cumulative distribution function, which is a monotonically increasing function, has to be calibrated. \citet{Schapire2002, Frank2009} use a heuristic approach to ensure this monotonicity requirement. Alternatively, we use the `isotonic' regression technique to calibrate the class probabilities. Isotonic regression is synonymous for monotonically increasing regression and is a technique for which efficient implementations are available \citep{ExplanationIsotonic}.
\begin{figure}
   \centering
 \includegraphics[scale=0.4]{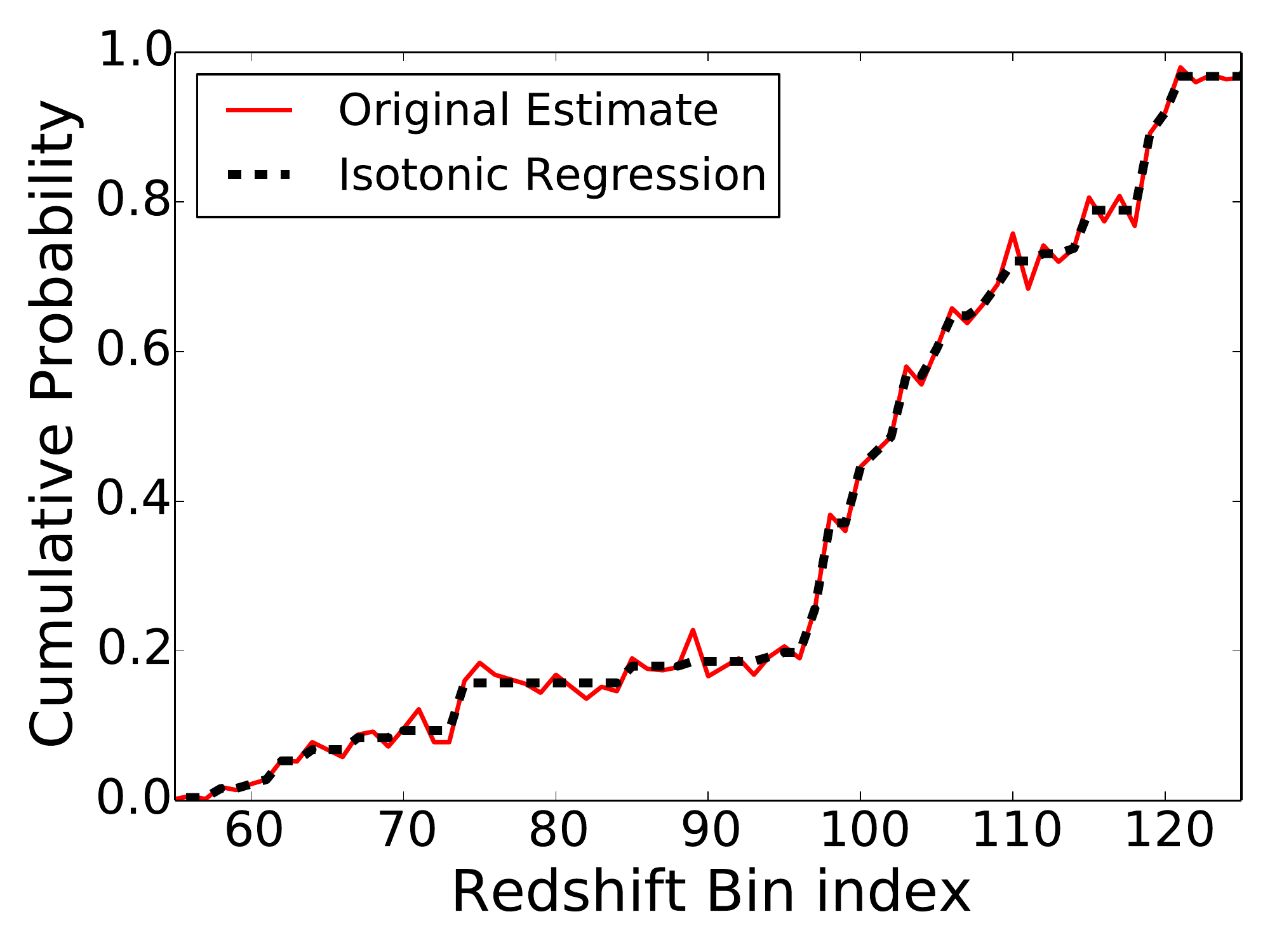}
   \caption{ \label{isoreg}  Ordinal classification can result in non monotonic cumulative distribution functions. We calibrate them using isotonic regression. Isotonic Regression (black) approximates the original estimate (red) as a monotonically increasing step function. } 
\end{figure}
For increasing bin index, isotonic regression optimizes the mean squared error between the original function values and the isotonic fit such that the fit is a monotonic increasing step function as shown in Fig. \ref{isoreg}.

We use bins of fixed size $\Delta z =  0.01$ in the range between the minimum and the maximum spectroscopic redshift value in the training set, since we found that equal frequency binning degrades photometric redshift accuracy for catalogues with long-tailed sample PDF. The weights sum to unity and are calculated using
\begin{equation}
 w_i(\mathbf{f}) = \frac{\hat{p}(b_i | \mathbf{f})}{n_{b_i}} \, ,
 \label{eq:weights_ocp}
\end{equation}
where $n_{b_i}$ is the number of training objects with a redshift value in bin $b_i$. The quantity $\hat{p}(b_i | \mathbf{f})$ is an estimate for the class probability that a newly queried object with photometry $\mathbf{f}$ has a spectroscopic redshift inside the bin $b_i$. The method used to obtain the class probability estimates $\hat{p}(b_i | \mathbf{f})$ is interchangeable (e.g., using Neural Networks \citealt{bonnet}, or the Random Forest \citealt{Frank2009, tpz}). 
We use the Random Forest algorithm for consistency with the Quantile Regression Forest and implemented the OCP algorithm using the `randomForest' \citep{randForest2002} package for the R programming language \citep{Rcitation}. 

The original paper by \citet{Schapire2002} used the histogram estimator defined in \citet{Frank2009} as
\begin{equation}
	\hat{p}(z | \mathbf{f}) = \sum_{i = 1}^{N_{\rm tr}} w_i(\mathbf{f}) \frac{I(b_{z_i} = b_{z})}{r_{b_z}} \, .
	\label{eq:histo_kern}
\end{equation}
Here $b_{z}$ is defined as an index denoting the bin in which $z$ is located and $r_{b_z}$ denotes the corresponding bin width. 
We can interpret this histogram as a weighted kernel density estimate with value $r_{b_z}^{-1}$ for all training set objects in a bin specified by $b_{z}$ and zero outside. \citet{Frank2009} improved the algorithm using a Gaussian kernel function and demonstrated its superiority over the histogram kernel in numerical experiments on machine learning benchmark datasets that are unrelated to the photometric redshift problem. 
\subsection{Bandwidth Selection}
\label{subsec:bandwidth_select}
The algorithms we use to obtain PDFs for individual objects require the selection of an appropriate bandwidth for the weighted kernel density estimator (Eq. \ref{eq:cond_prob}). This section proposes a bandwidth selection scheme that selects the bandwidth for the Gaussian kernel during model tuning using the MNLL. 

The choice of a proper bandwidth depends on the shape of the underlying distribution and the number of objects available to construct the estimator. Assuming a normal distribution and a Gaussian kernel function one can obtain the optimal bandwidth as  
\begin{equation} 
	\sigma_{\rm scott} = 1.06 \frac{\hat{\sigma}}{N^{1/5}} \, ,
	\label{eq:stand_scott}
\end{equation}
where $\hat{\sigma}$ is the sample standard deviation and $N$ denotes the number of objects. This so-called `Scott's rule' is commonly used in the machine learning and statistics literature \citep[e.g.][]{Takeuchi2007, Wang2007}. To apply this bandwidth selection rule to weighted data, we need to calculate the weighted standard deviation from the weighted training set using Eq. \ref{eq:standdev}. 
Scott's rule gives a good first estimate of a suitable bandwidth for distributions which are approximately normal. 

Photometric redshift PDFs are in general not Normal distributions and Eq. \ref{eq:stand_scott} can pick a non-optimal bandwidth.
Thus, we modify Eq. \ref{eq:stand_scott}
\begin{equation}
\sigma_{\rm scott} = a \, \frac{\hat{\sigma}}{N^{1/5}} \, ,
\label{eq:bw_select_scott}
\end{equation}
with a pre-factor $a$ that is chosen to minimize the MNLL on the validation set.
We can stack the $N_{\rm te}$ individual PDFs in the test set using an individual bandwidth $\sigma_{a}$ for each object 
\begin{equation}
 \hat{p}(z) = \sum_{a = 1}^{N_{\rm te}}  w_{{\rm stack}, a} \sum_{i = 1}^{N_{\rm tr}} 
w_{i}(\mathbf{f}_{a}) \mathcal{N}(z, z_i, \sigma_{a}) \, ,
\label{eq:pz_stacked}
\end{equation}
or we can use a global bandwidth $\sigma_{a} = \sigma$.
\subsection{The Gaussian Mixture Model Estimator}
\label{subsec:fit_gmm}
Storing the individual PDFs obtained by weighted kernel density estimation for every element in the test set requires a large amount of storage. \citet{Sparse2014} proposed several different methods, including a Gaussian mixture model, to more efficiently store a previously obtained estimate. The authors store individual PDFs using 10 - 20 numbers compared with 200 used previously. Instead of giving a previously estimated individual PDF a sparse representation, we fit the Gaussian mixture model directly to the weighted spectroscopic redshift values in the training set and ensure model sparsity by penalizing the model likelihood dependent on the number of components in the mixture model. 

More specifically, we fit the Gaussian mixture to the weighted spectroscopic data with the expectation maximization algorithm \citep[for an introduction see][]{chen2010demystified} as implemented in the Rmixmod package \citep{Biernacki2006, Rmixmod}. In \S \ref{subsub:comp_annz} and \S \ref{subsec:smnlp} during the analysis using CFHTLS, we select the number of Gaussian components for each object in the test set using the normalized entropy criterion \citep{Celeux1996, Biernacki1999}, appreviated as NEC in the following. The maximum number of Gaussian components that can be included in the mixture model is a parameter that is selected during model tuning as described in \S \ref{eq:testingframe}.

For a \textit{K}-component Gaussian mixture model fitted on the weighted training data, the NEC criterion reads
\begin{equation}
	NEC(K) = \frac{E(K)}{L(K) - L(1)}
	\label{eq:nec_crit}
\end{equation} 
where $L(K)$ denotes the maximum weighted log-likelihood
\begin{equation}
 L(K) = \sum_{i = 1}^{N} w(\mathbf{f}_i) \log{\left(\sum_{k = 1}^{K} \alpha_{k} \mathcal{N}\left(z_{{\rm spec}, i}, \mu_{i}, \sigma_{i}\right)\right)}
\end{equation}
 for the $K$ component Gaussian mixture model. The entropy $E(K)$ is defined as
\begin{equation}
E(K) = - \sum_{k = 1}^{K} \sum_{i = 1}^{N} w(\mathbf{f}_i) \gamma_{k}\left(z_{{\rm spec}, i}\right) \log{\left(\gamma_{k}\left(z_{{\rm spec}, i}\right)\right)} \ \leq 0 \, ,
\end{equation}
where the definition of the component weight proportions, following Eq. \ref{eq:response}, is used.
We pick the number of components $K$ such that the NEC criterion is minimized, where $NEC(1) = 1$ \citep{Biernacki1999}. 

The NEC criterion normalizes the entropy by the maximum weighted log-likelihood, in which the offset for a one-component mixture is substracted. There are two reasons \citep{Celeux1996} why we cannot use the entropy $E(K)$ directly. The entropy for $K = 1$ provides a lower bound 
\begin{equation}
E(K) \geq E(1) \ \ \ \forall K > 1
\end{equation}
and the maximum weighted log-likelihood function is an increasing function of $K$, which makes $E(K)$ unequal for different values of $K$. The entropy term $E(K)$ measures how much overlap there is between the different components of the Gaussian mixture model. In the case where the components in the model fit completely separated data clusters, the entropy term approaches zero. If we select too many components, the quantity $E(K)$ will increase because the components will overlap strongly. This can be compensated by the higher likelihood of the more complex model. In this way, we can efficiently determine a suitable number of components to include into the mixture.
\subsection{Highest Weight Element}
\label{subsec:hwe}
A common application for individual PDFs is the estimation of the sample PDF. Storing and processing individual PDFs is computationally expensive. We propose the Highest Weight Element (hereafter HWE), a single point estimate for each object from which we can accurately reconstruct the sample PDF. We first run the QRF algorithm to determine weights as for individual PDF estimation. Instead of using the individual PDF, we select the spectroscopic redshift value that is associated with the maximum weight. If more than one spectroscopic redshift value has the same maximum weight, we randomly select one of those values.

\section{DATASET}
\label{dataset}
We use photometric imaging data from the CFHTLS Wide survey using the following bands $u^*$, $g'$, $r'$, $i'$ and $z'$-band as obtained from the public CFHTLenS data release \citep{Erben2012} \footnote{http://www.cfhtlens.org}. We obtain the photometry analogously to \citet{Brimioulle2013}, i.e., we degrade all images to match the band with the worst seeing, and use the unconvolved $i'$-band as the detection band and the convolved frames as the extraction band. Then we correct for the remaining zeropoint calibration uncertainties and varying galactic extinction by comparing the measured star colours from the catalogues with predictions of the Pickles star library \citep{Pickles1998}. In this way, we eliminate possibly remaining field-to-field variations in the photometric calibration.

We then match our photometric catalogues to public spectroscopic redshift samples. These samples are the Visible Multiobject Spectrograph (VIMOS) VLT Deep Survey (VVDS) \citep{Fevre2004, Garilli2008}, VVDS-F22, the Deep Extragalactic Evolutionary Probe-2 (DEEP-2) program \citep{Davis2007, Vogt2005, Weiner2005} and the VIMOS Public Extragalactic Redshift Survey (VIPERS) \citep{Garilli2014,Guzzo2014}. We only make use of spectroscopic redshifts with confidence values of at least 95\% and only use pointings where the i'-data is available and where the i'-band serves as detection band.

This produces a total sample of 28159 objects with $i' \le 22.5$ and additional 5893 objects with $22.5 < i' \le 24.5$ with spectroscopic redshifts and five band photometry.
\begin{figure}
   \centering
 \includegraphics[scale=0.4]{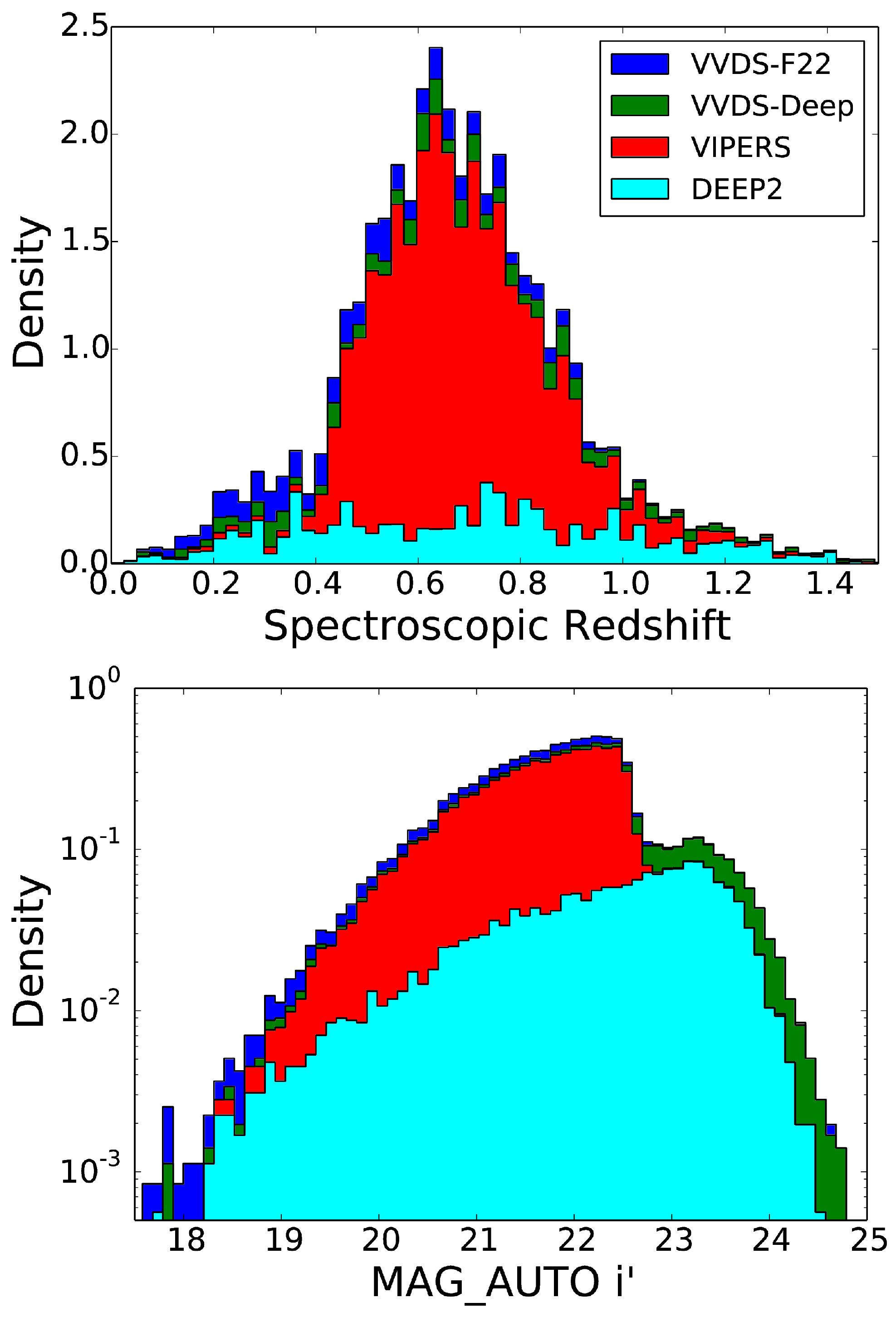}
   \caption{ \label{zspec_plot} Spectroscopic redshift and \texttt{MAG\_AUTO} $i'$ distributions of the compiled dataset described in \S \ref{dataset}. Objects matched from different spectroscopic surveys are indicated by different colors. We limit the spectroscopic redshift range to $z_{\rm spec} < 1.5$ in the plots excluding 34 objects with higher redshift. } 
\end{figure}

\section{RESULTS}
\label{results}
Future large area photometric surveys will produce large amounts of photometric data for which we need to obtain redshift information. Efficiency in terms of runtime and disk space will be important in order to use algorithms for photometric redshift estimation effectively on these large datasets. Additionally we are required to produce high quality photometric redshift PDFs in order to obtain accurate constraints on, for example, cosmological parameters or cluster masses.

We use the public CFHTLS data described in \S \ref{dataset}, to compare the accuracy of photometric redshift PDFs estimated by the algorithms described in \S \ref{algorithms}. We show that these methods improve the modelling of angular correlation functions, cluster mass estimates, and the modelling of shear correlation functions compared to results obtained with the Neural Network code ANNz \citep{collister} commonly used in the literature \citep[e.g.,][]{Sheldon2009, SPT2011, Smith2012useANNz, Planck2015}.
\subsection{Comparison with ANNz}
\label{subsub:comp_annz}
We train an ensemble of 20 Neural Networks with two hidden layers, each consisting of 12 nodes, following the methodology described in \S \ref{eq:testingframe}. 

The photometric redshift estimates obtained from ANNz are competitive compared to those obtained with the template fitting code PhotoZ \citep{Bender2001, brimouille2008, Greisel2013} in terms of common photometric redshift performance metrics. As shown in Table \ref{cap:annz_photoZ}, ANNz improves upon the photometric redshift performance obtained with PhotoZ by 46\%, 29\%, 88\% and 12\% in terms of outlier rate, scatter, bias and spread of the residuals. The outlier rate $\eta$ is defined as the fraction of objects with $\left|z_{\rm spec} - z_{\rm phot}\right| > 0.15$. The bias $\left\langle \Delta z \right\rangle$ and scatter $\sigma(\Delta z)$ are the mean and standard deviation of the distribution of the residuals $\Delta z = z_{\rm phot} - z_{\rm spec}$. The spread of the residual distribution is measured by the $\sigma_{\rm 68}$ metric which is defined as half the difference between the 16\% and 84\% quantile.
\begin{table}
\begin{center}
  \begin{tabular}{ l | l | l | l | l } 
       & $\eta$   &  $\sigma(\Delta z)$ & $\left\langle \Delta z \right\rangle$ & $\sigma_{68}$  \\ \hline\hline
ANNz   & 1.23\%   & 0.092               & -0.001                                & 0.044    \\ 
PhotoZ  & 2.27\%  & 0.129               & -0.008                                & 0.050   \\
\\
 \end{tabular}
\caption{ \label{cap:annz_photoZ} Point prediction performance of the Neural Network code ANNz and the template fitting code PhotoZ quantified by the metrics described in \S \ref{subsub:comp_annz}.  }
\end{center}
\end{table}
\begin{figure}
   \centering
 	\includegraphics[scale=0.4]{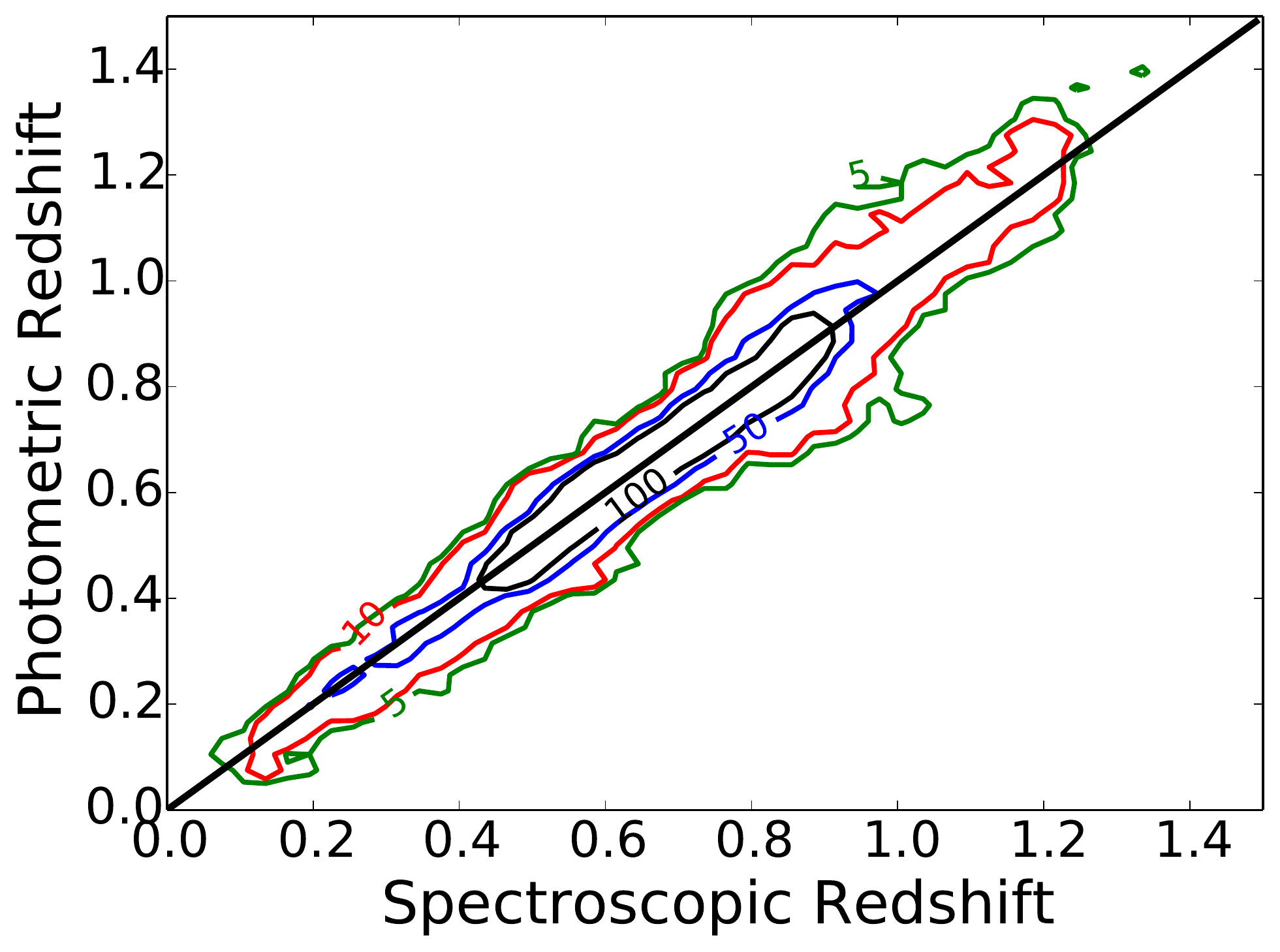}
   \caption{ \label{scatter_annz_zspec} Density contours of photometric redshift estimates from ANNz against the spectroscopic redshift.  } 
\end{figure}

\begin{figure}
   \centering
 	\includegraphics[scale=0.4]{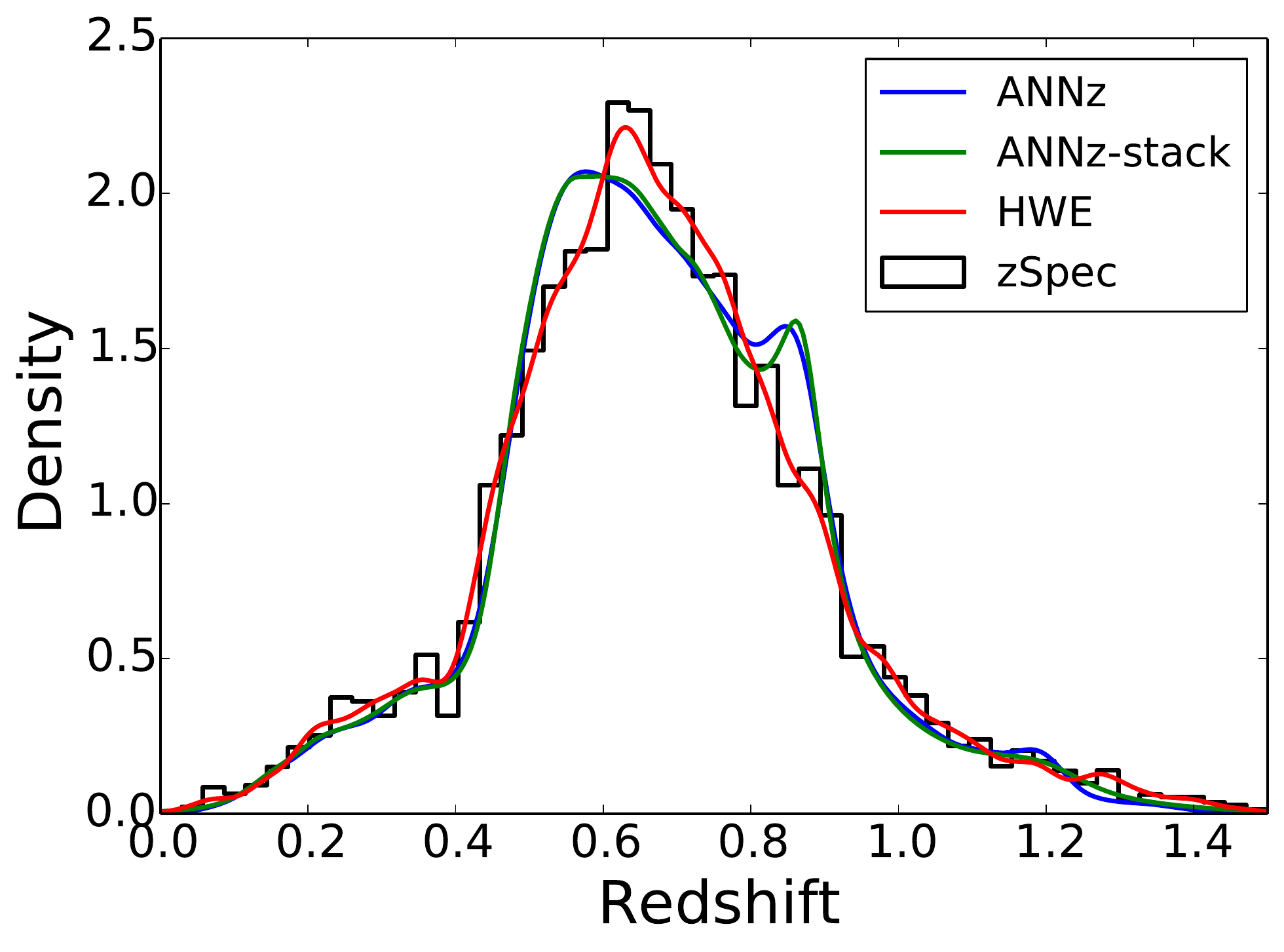}
   \caption{ \label{comparison_distri_annz_hwe} Sample PDF estimated using ANNz and the Highest Weight Element. The histogram shows the true spectroscopic redshift distribution. } 
\end{figure}

The quality of the photometric redshifts obtained with ANNz is illustrated in Fig. \ref{scatter_annz_zspec}. It shows a tightly aligned correlation between photometric and spectroscopic redshift.
We estimate sample PDFs from the ANNz point predictions and the stacked Normal densities constructed from the ANNz error estimates, in the following referred to as `ANNz-stack'.
While showing excellent point prediction performance, ANNz and ANNz-stack do not accurately estimate the sample PDF as shown in Fig. \ref{comparison_distri_annz_hwe}. The sample PDF constructed from ANNz-stack deviates from the true spectroscopic redshift distribution in the central redshift range $[0.45, 0.85]$. 
We will show in the following sections that these deviations from the true spectroscopic redshift PDF introduce a systematic bias in several important analyses in cosmology. 
\begin{table}
\begin{center}
  \begin{tabular}{|l||l|l|l|l|} 
        &  Total	       & $[0, 0.585[$ & $[0.585, 0.7488[$ & $[0.7488, 3.818[$    \\\hline\hline
OCP     &  -1.3577   &   -1.3905  &    -1.6432      &     -1.0395        \\
NOCP    &  -1.2847   &  -1.3029   &  -1.5880        &   -0.9648          \\
QRF     &  -1.3483   & -1.3627    &    -1.6470      &     -1.0347         \\
GMM     &  -1.3181   & -1.3591    &    -1.5606      &     -1.0354          \\  
ANNz    &  -1.1588    & -1.3138    &    -1.4891      &   -0.6731           \\
\\
 \end{tabular}
\caption{\label{cap:mnlp_performance} MNLL of the Quantile Regression Forest (QRF), the classification based PDF estimation algorithms (OCP/NOCP) and the OCP algorithm used with the Gaussian mixture model (GMM). The values are evaluated over the full spectroscopic redshift range and in three bins. The result is illustrated in Fig. \ref{comparison_mnlp}. }
\end{center}
\end{table}
\begin{figure}
   \centering
 \includegraphics[scale=0.4]{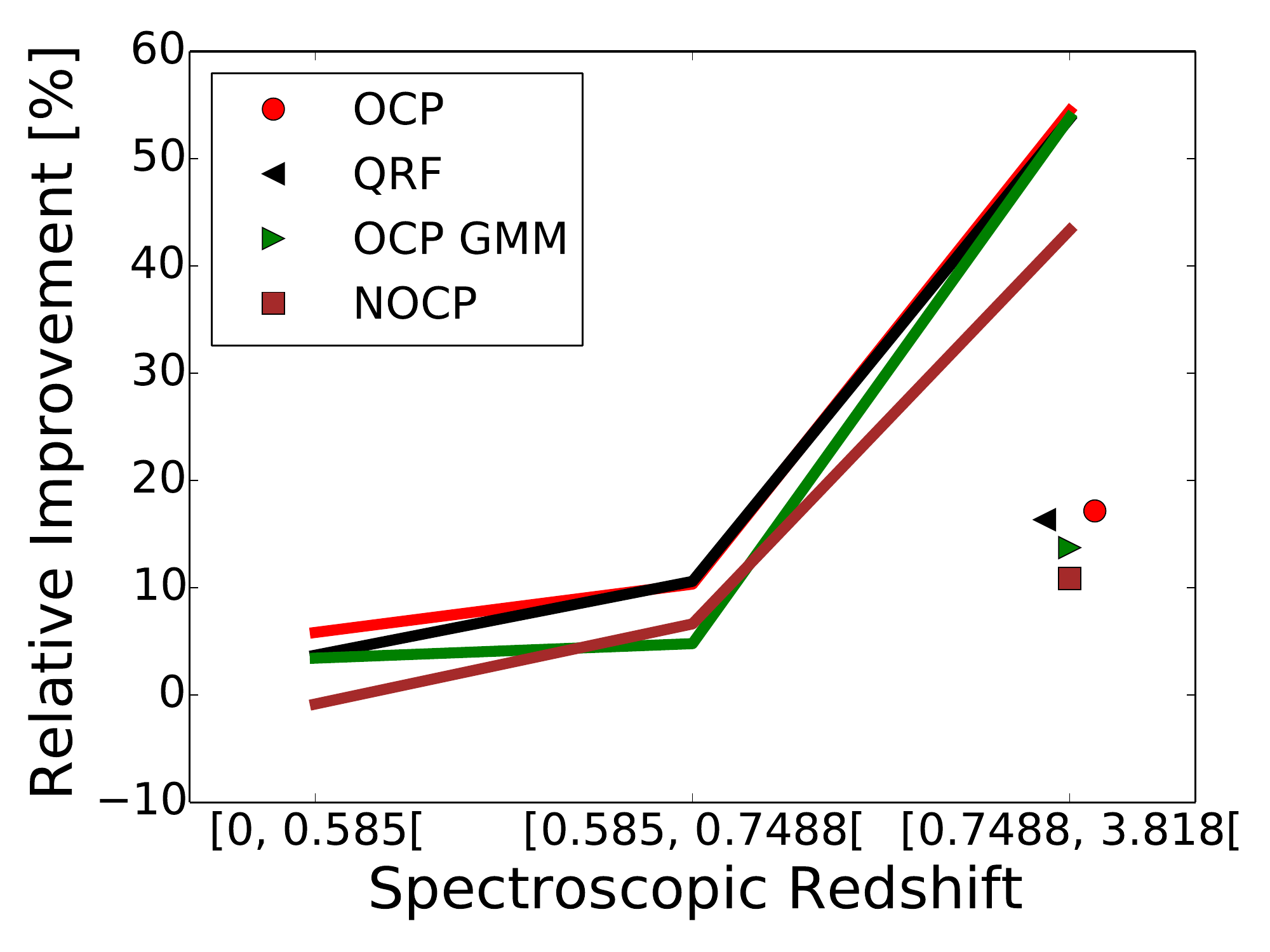}
   \caption{ \label{comparison_mnlp} Relative improvement in MNLL over the performance of ANNz-stack. We compare the classification-based PDF estimators (OCP, NOCP), the ordinal classification PDF estimator combined with a Gaussian mixture model (OCP GMM) and the Quantile Regression Forest (QRF) in three spectroscopic redshift bins. The plotted points show the average improvement over the full spectroscopic redshift range.  } 
\end{figure}
To compare the quality of photometric redshift PDFs of individual objects (individual PDFs), we evaluate the MNLL (Eq. \ref{eq:mnlp_orig}) of the four discussed algorithms (QRF, NOCP, OCP and OCP GMM) on the full range of redshift values and in three redshift bins ($[0, 0.585[$, $[0.585, 0.7488[$ and $[0.7488, 3.818[$). The results are shown in Table \ref{cap:mnlp_performance} and illustrated in Fig. \ref{comparison_mnlp}. QRF, NOCP and OCP employ the weighted kernel density estimate. OCP GMM denotes the Gaussian mixture model applied in combination with weights determined using the ordinal classification method described in \S \ref{subsec:ocp}. 

We illustrate the relative improvement ${\rm MNLL}_{\rm rel}$ gained by applying these algorithms compared with ANNz-stack
\begin{equation}
	{\rm MNLL}_{\rm rel} = \left(\frac{{\rm MNLL}_{\rm ANNz} - {\rm MNLL}_{\rm alg.}}{|{\rm MNLL}_{\rm ANNz}|}\right)
\end{equation}
in Fig. \ref{comparison_mnlp}. A high value in terms of ${\rm MNLL}_{\rm rel}$ translates into an improvement in the log-likelihood of the individual PDFs over those obtained with ANNz-stack. The boundaries of the redshift intervals are picked such that they contain approximately the same number of test set objects. All discussed methods improve over ANNz-stack. For the highest redshift objects, our methods show improvement of up to 50\%. The OCP routine performs the best and improves the NOCP routine. This verifies the superiority of the ordinal classification technique. The Quantile Regression Forest performs on par with OCP. OCP GMM shows mediocre results, but provides the most efficient parametrization using a single Normal density per object.  
\subsection{Stacked photometric redshift distribution}
\label{subsec:smnlp}
Applications like shear tomography require the photometric selection of objects in a certain redshift range. We stack the individual PDFs compared in \S \ref{subsub:comp_annz} using weights that quantify their overlap with a certain redshift interval using Eq. \ref{eq:stack_weights}. These estimates are compared with the weighted kernel density estimate obtained from the spectroscopic redshift values using the same weights. The weights are determined using each of the OCP, NOCP, OCP GMM and QRF methods individually. The Highest Weight Element (HWE) uses the weighted kernel density estimate with weights determined using the QRF algorithm. We use Scott's rule to choose the bandwidth for the weighted kernel density estimates of the HWE and the spectroscopic redshift values.
The sample PDFs obtained with the OCP, NOCP and QRF algorithms are very similar. We therefore restrict the following discussion to the OCP method.

\begin{figure*}
   \centering
 	\includegraphics[scale=0.8]{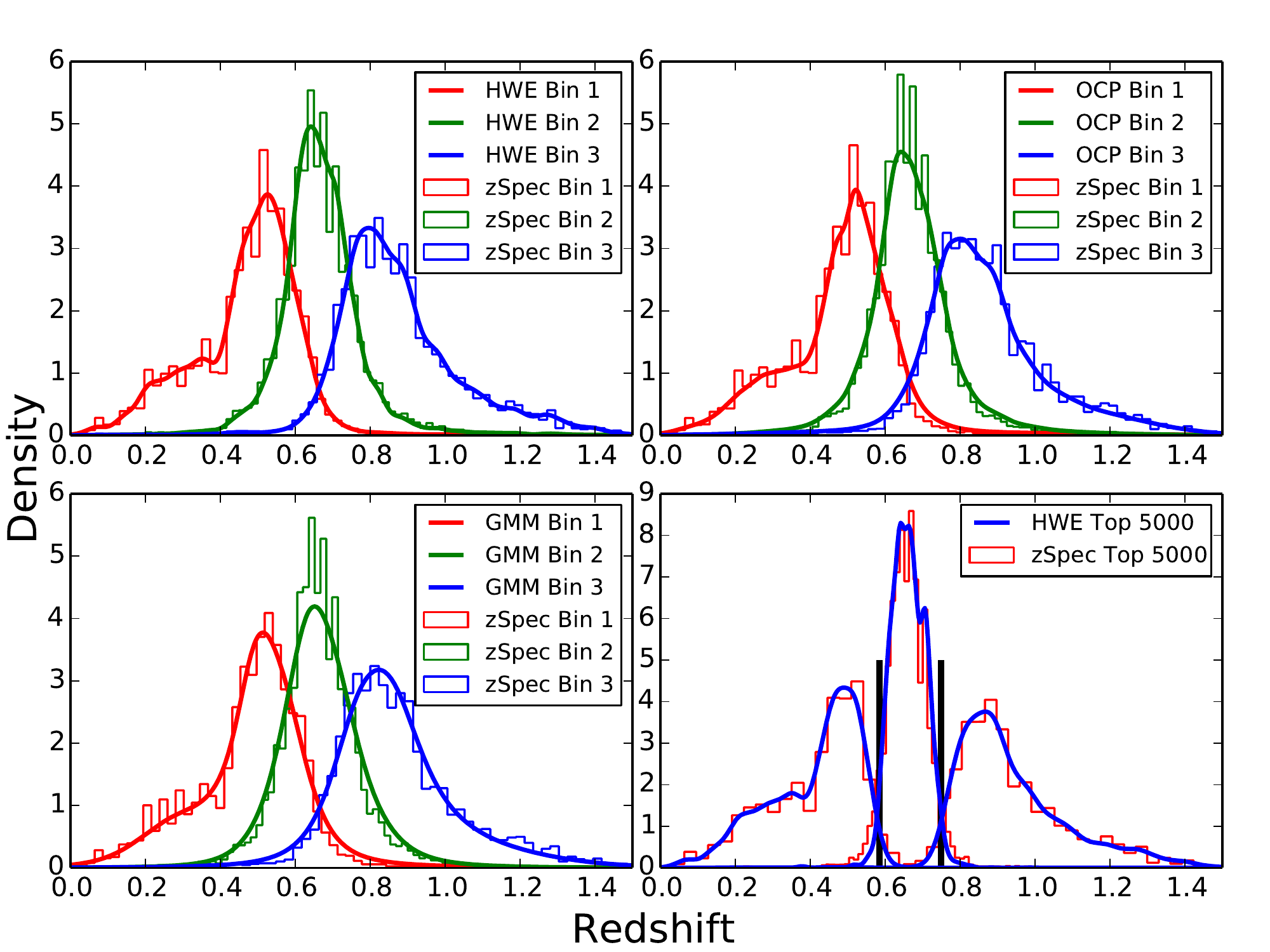}
   \caption{ \label{comparison_distri_three_bins} Sample PDFs weighted in three redshift intervals $[0, 0.585[$, $[0.585, 0.7488[$ and $[0.7488, 3.818[$. The PDFs are obtained using the Highest Weight Element (upper left), the ordinal classification PDF estimator (upper right) and the ordinal classification PDF estimator combined with a Gaussian mixture model (lower left). The histograms show the weighted spectroscopic redshift distribution using weights determined using the respective algorithms. The lower right panel shows the weighted distribution of the HWE predictions for the objects with the 5000 highest weights in the three intervals (blue) and the corresponding weighted histogram of spectroscopic redshifts (red).    } 
\end{figure*}

The results shown in Fig. \ref{comparison_distri_three_bins} compare the weighted sample PDFs obtained with the HWE, OCP and OCP GMM methods in the redshift intervals $[0, 0.585[$, $[0.585, 0.7488[$ and $[0.7488, 3.818[$. They differ mainly in the amount of smoothing present in the estimate. Notably the OCP GMM method oversmooths features in the density estimate. This is because a single Gaussian was selected during model tuning based on the performance of individual object PDFs. Allowing more components reduces the amount of smoothing. The HWE is competitive with methods that estimate the individual PDFs, with the advantage that the HWE is extremely efficient to calculate and, being a point estimate, requires storing only a single floating point number per object. 
\begin{figure*}
   \centering
 	\includegraphics[scale=0.8]{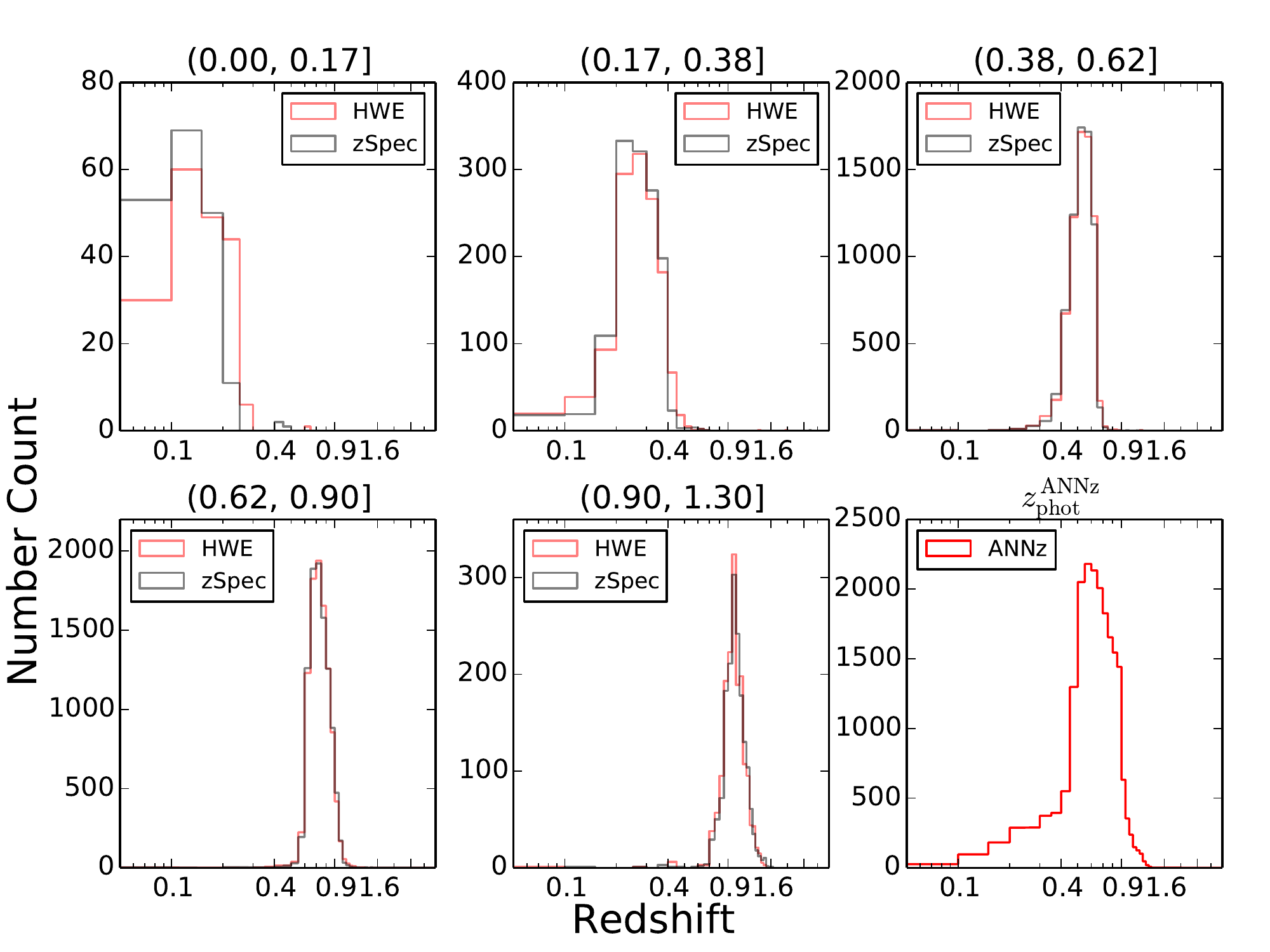}
   \caption{ \label{comparison_joachimi} Sample PDFs estimated using the HWE for subsamples selected in analogy to \citet[][Fig. 1]{Benjamin2013} using a cut at \texttt{MAG\_AUTO} $i' < 23.0$. The subsamples are selected using the photometric redshift estimates from ANNz in intervals shown in the subfigure titles.  } 
\end{figure*}

The weighted distributions of all methods have tails that extend outside the desired redshift range.
We can reduce these tails by stacking only the objects with the highest weight in the respective redshift bin as demonstrated in the lower right panel of Fig. \ref{comparison_distri_three_bins}. We estimate the sample PDF from the HWE predictions of the objects with the 5000 highest weights in the respective redshift bin. The estimated weighted sample PDF of these objects has less overlap with neighbouring redshift bins, compared with the estimate that incorporates all objects. Furthermore it agrees well with the equally weighted spectroscopic redshift distribution of the corresponding objects.

Instead of weighting the objects in the respective redshift range, we can select objects based on a photometric redshift point estimate in analogy to \citet{Benjamin2013}. We perform the same cut in \texttt{MAG\_AUTO} $i' < 23.0$ and estimate the sample PDF in the same photometric redshift intervals selected after our ANNz estimate. The results for the HWE are shown in Fig. \ref{comparison_joachimi} and agree well with the spectroscopic redshift distribution.
The agreement is better in the central bins, which contain more objects, because the histogram approximates the underlying distribution better.
\subsection{Applications to Cosmology}
\label{subsec:application}
We now investigate how the previously discussed methods can be used to improve analyses that use photometric redshifts. We estimate the sample PDF using the Highest Weight Element and ANNz. We use kernel density estimates with bandwidths selected using Scott's rule.

Where required, we impose a flat $\Lambda$-CDM cosmology with $\Omega_{m} = 0.3$, $\Omega_{\Lambda} = 0.7$, $n_{s} = 0.96$, $H = 0.7$, $\sigma_{8} = 0.79$.
\subsubsection{The Angular Power Spectrum}
\label{subsub:the_angular_power}
The angular power spectrum measures the clustering of galaxies and is an important tool to constrain cosmological models. 

In the following we adopt the notation of \citet{Thomas2010}. Consider the line-of-sight projection of the 3D mass distribution in the universe, $\delta_{2\mathrm{D}}$. The harmonic modes of $\delta_{2\mathrm{D}}$ are given by
\begin{equation}
	\delta_{\ell} = i^{\ell} \int \frac{d^3 k}{(2 \pi)^3} \delta(\mathbf{k}) W_\ell(k) \ ,
\end{equation}
where the window function $W_\ell(k)$ is sensitive to the sample PDF of light sources, $p(z)$, and can be computed by the integral 
\begin{equation}
\label{eq:angular_power_spectrum_window_function}
	W_\ell(k) = \int p(z) D(z) \left(\frac{d z}{d x}\right) j_\ell(k z) dz \ .
\end{equation}
Here $D(z)$ is the linear growth factor, $j_\ell(k z)$ are the Bessel functions and $\left(\frac{d z}{d x}\right)$ relates the redshift to the radial comoving coordinate $x$.

The angular power spectrum $C_\ell$, is the variance of the modes $\delta_\ell$\footnote{In our analysis, we are assuming a galaxy-dark matter bias equal to one.},
\begin{equation}
	C_{\ell} = \left\langle \delta_\ell \delta_\ell^{*} \right\rangle = 4 \pi \int \Delta^{2}(k) W_{\ell}^{2} (k) \frac{dk}{k} \, ,
\end{equation}
where the dimensionless 3D power spectrum $\Delta^{2}(k)$ is given in terms of the usual 3D matter power spectrum $P_\delta(k)$ as
\begin{equation}
	\Delta^{2}(k) = \frac{4 \pi k^{3} P_\delta(k)}{(2\pi)^3} \ .
\end{equation}
From Eq. \ref{eq:angular_power_spectrum_window_function} is can be seen, that the modelling of $C_\ell$ depends highly on the assumed sample PDF of the data. We use the distributions shown in Fig. \ref{comparison_distri_annz_hwe} to model the angular correlation power spectrum with the CLASS software package \citep{Blas2011}. 
\begin{figure}
   \centering
 \includegraphics[scale=0.4]{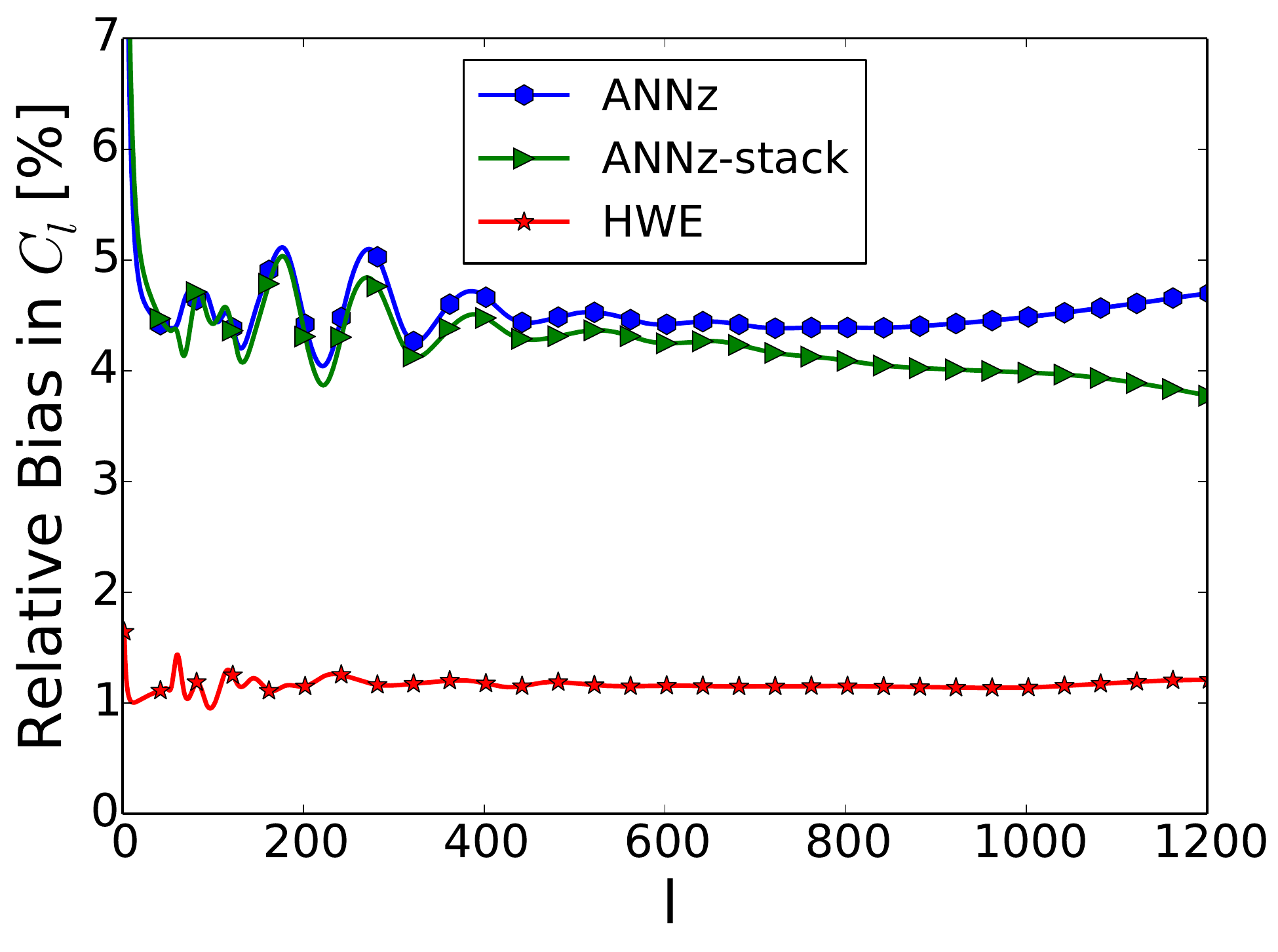}
   \caption{ \label{comparison_autocorrelation} Bias in the angular correlation power spectrum obtained for different estimates for the sample PDF. We restrict the comparison to $\ell < 1200$.   } 
\end{figure}
We define the bias introduced by the $C_{\ell}^{\rm phot}$ of the angular correlation function estimated using photometric redshifts, as the relative difference to the results based on the PDF of spectroscopic redshifts $C_{\ell}^{\rm spec}$:
\begin{equation}
	{\rm Bias}_{\rm C_\ell} = \left(\frac{C_{\ell}^{\rm phot} - C_{\ell}^{\rm spec}}{C_{\ell}^{\rm spec}} \right) \, .
\end{equation}
The resulting biases are shown in Fig. \ref{comparison_autocorrelation}. We find that the results obtained with the Highest Weight Element have a lower systematic bias in $C_{\ell}$ by a factor of four compared to the ANNz results and that the improvement is almost independent of $\ell$.
\subsubsection{Gravitational Lensing}
\label{subsub:lensing_app}
We investigate two important applications in gravitational lensing: quantifying cluster masses by the light deflection from background sources, and obtaining cosmic shear correlation functions. In contrast to the previously considered analysis of the angular correlation function, applications in gravitational lensing require careful selection of sources with successfully measured shapes. Since the spectrophotometric dataset used previously is not representative for datasets generally used in gravitational lensing analyses, we first weight our catalogue such that it mimics a CFHTLS shape catalogue. To do this, we obtain a photometric shape catalogue from public CFHTLS data, which is then used as the reference to weight the spectrophotometric dataset.  
\subsubsection{Catalogue Creation and Weighting}
\label{subsubsec:catalogue_creation}
Whether the shape of an object can be measured, depends primarily on its intrinsic size and magnitude in the respective band. We therefore reweight our spectrophotometric catalogue such that it resembles a CFHTLS shape catalogue in terms of these properties. We obtain the shape catalogue in analogy to \citet{Brimioulle2013} for the full CFHTLS survey region. Intrinsic sizes $s_{\rm intr}$ are calculated for each object from the measured ${\rm FWHM_{image}}$ and corrected for seeing as follows 
\begin{equation}
 s_{\rm intr} = \sqrt{{\rm FWHM_{image}}^2 - \left\langle{\rm FWHM_{\rm psf}}\right\rangle^2} \, ,
 \label{eq:intrins_size}
\end{equation}
where $\left\langle{\rm FWHM_{\rm psf}}\right\rangle$ is the average size of the point spread function for the respective chip\footnote{We work on image stacks, but (as in \citealt{Brimioulle2013}) only consider objects, for which all images contribute to the stack from the same CCD-chip.}.

In this way, we obtain $s_{\rm intr}$ and \texttt{MAG\_AUTO} $i'$ entries for each object in the shape and spectrophotometric catalogue. We now determine weights for the spectrophotometric catalogue such that, after weighting, it matches the size and magnitude distribution of the shape catalogue.
Furthermore, the results obtained with the reweighted spectrophotometric catalogue have to be robust against the removal of the objects with the highest weights \citep{dessanchez}. 
Since we do not have enough spectroscopically observed objects to mimic the shape catalogue at the faint end, we have to employ a magnitude cut in order to fulfill both requirements. For the analyses presented in \S \ref{subsubsec:cluster_mass} and \S \ref{subsubsec:cosmic_shear}, we employ a magnitude cut at \texttt{MAG\_AUTO} $i' < 23.5$ and \texttt{MAG\_AUTO} $i' < 23.0$, respectively. We give a detailed discussion of these cuts in the appendix.

\begin{figure}
   \centering
 \includegraphics[scale=0.4]{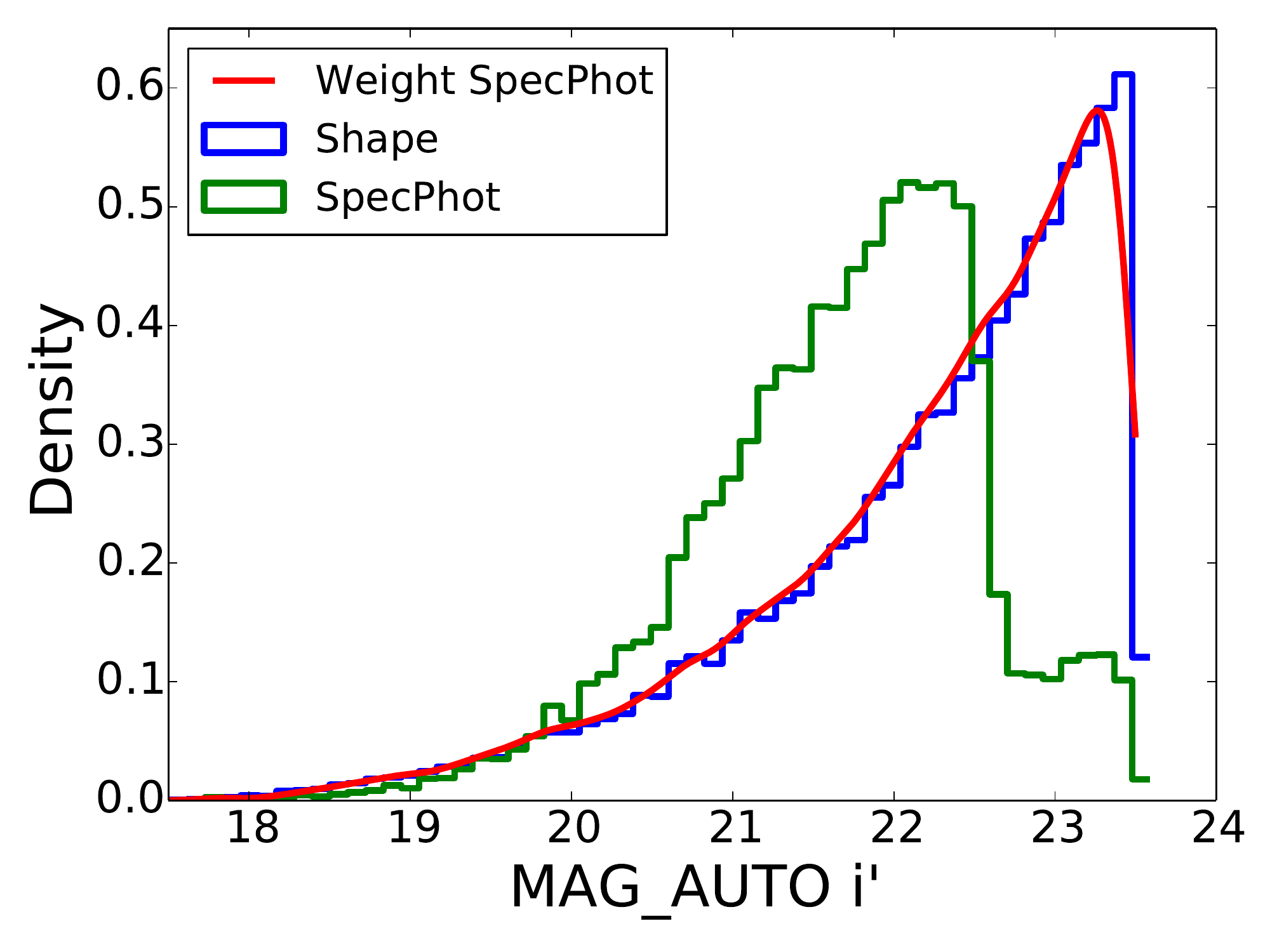}
   \caption{ \label{magauto_weight} Distributions in \texttt{MAG\_AUTO} $i'$ band for the original spectrophotometric dataset, the re-weighted spectrophotometric dataset and the shape catalogue for \texttt{MAG\_AUTO} $i' < 23.5$. } 
\end{figure}

\begin{figure}
   \centering
 \includegraphics[scale=0.4]{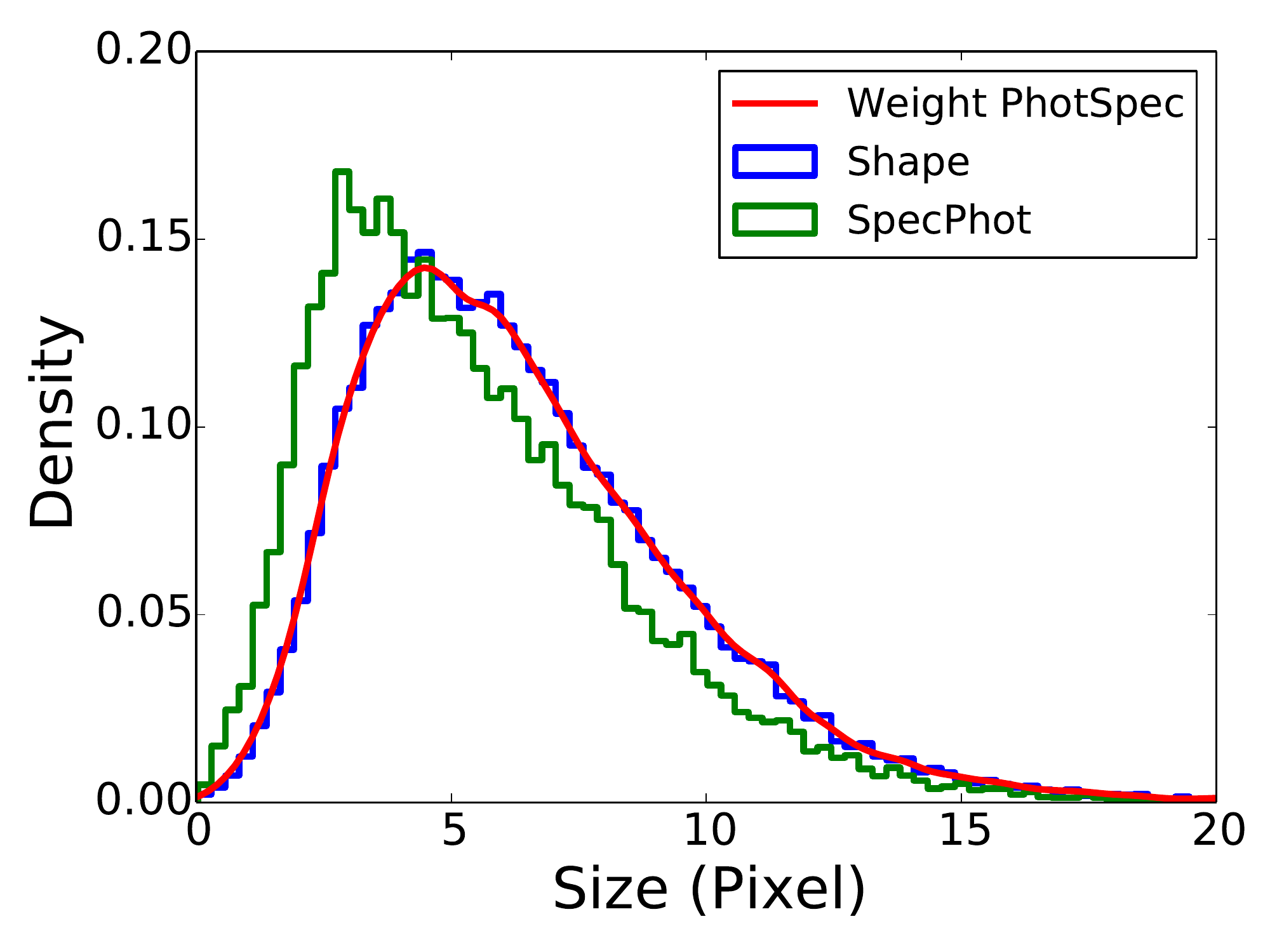}
   \caption{ \label{fwhm_weight} Distributions in intrinsic size (Eq. \ref{eq:intrins_size}) for the original spectrophotometric dataset, the re-weighted spectrophotometric dataset and the shape catalogue for \texttt{MAG\_AUTO} $i' < 23.5$. } 
\end{figure}

\begin{figure}
   \centering
 	\includegraphics[scale=0.4]{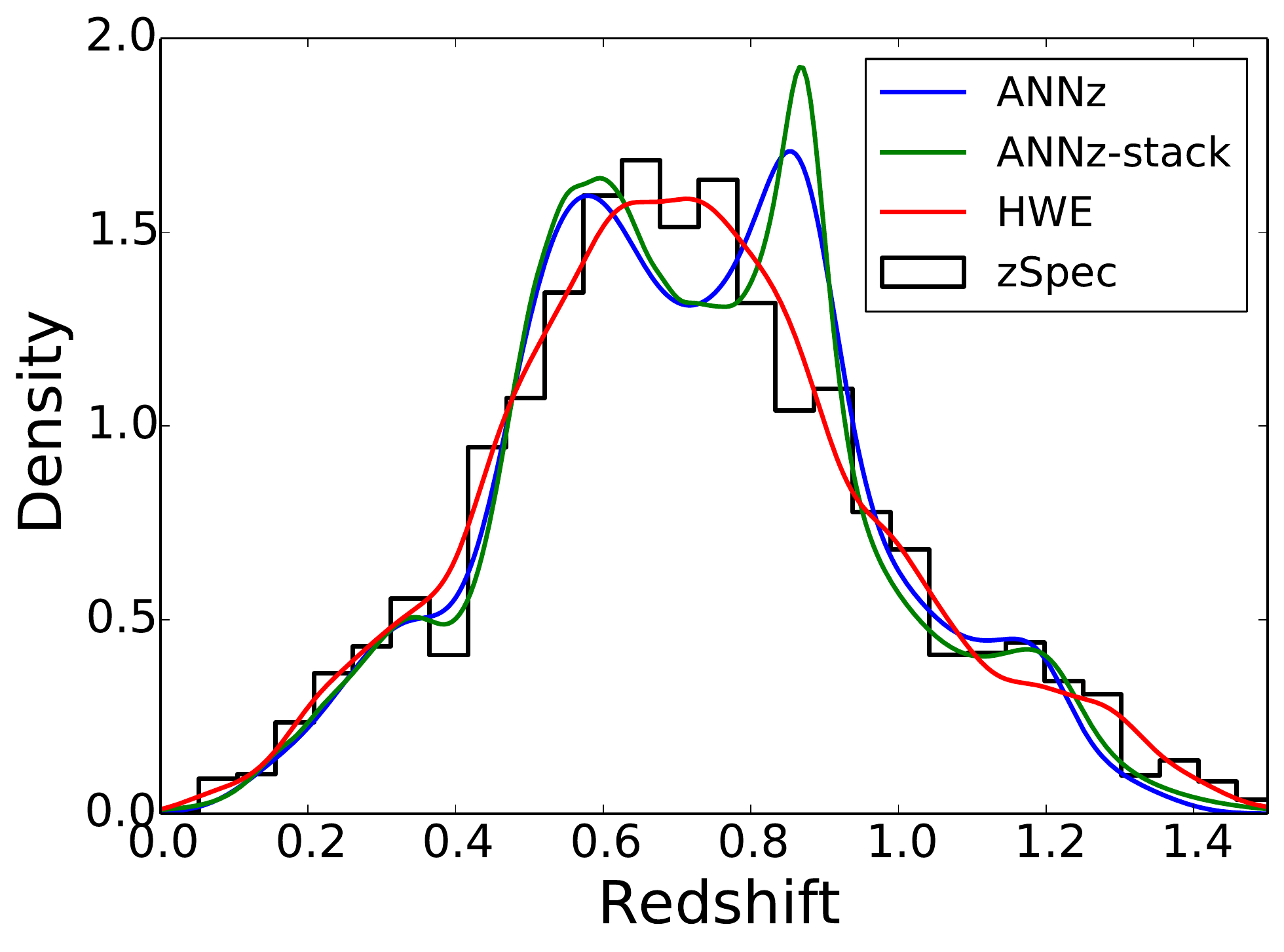}
   \caption{ \label{weighted_comparison_distri_annz_hwe} Weighted stacked sample PDF estimated using ANNz, ANNz-stack and the Highest Weight Element. The histogram shows the weighted spectroscopic redshift distribution. We use a cut on \texttt{MAG\_AUTO} $i' < 23.5$. The used weights and cuts are described in \S \ref{subsubsec:catalogue_creation}.   } 
\end{figure}

We combine bootstrap re-sampling with the $k$ nearest neighbour estimator to determine weights for the elements in the spectrophotometric catalogue, such that the weighted catalogue mimics the distribution of the shape catalogue in the two dimensional space spanned by the intrinsic size of the objects and their magnitude \texttt{MAG\_AUTO} $i'$. To this end, we draw bootstrap samples from the shear catalogue and find the $k$ nearest neighbours in the spectrophotometric catalogue. The nearest neighbour of an object in the spectrophotometric catalogue is the object in the shear catalogue with the lowest Euclidean distance to this object. Accordingly, the $k$ nearest neighbour algorithm selects the $k$ nearest objects with respect to the Euclidean distance. The number of times an object in the spectrophotometric catalogue is selected as one of the $k$ nearest neighbours corresponds to its weight. This process is similar to previous work done by \citet{Lima2008}, which employs a nearest neighbour based approach to determine weights for objects in a spectroscopic sample to estimate the sample PDF of the photometric data. In contrast to our method, which is based on bootstrap re-sampling, they calculate the density ratio between the distributions characterizing the two catalogues using a nearest neighbour approach. For the data at hand, we draw $10^{6}$ bootstrap samples and consider three nearest neighbours $k = 3$. This method accurately weights the spectrophotometric data to mimic the size and i-band magnitude distributions of the shape catalogue, as shown in Figs. \ref{magauto_weight} and \ref{fwhm_weight}. The following analysis uses the estimated weights to weight the sample PDF of ANNz, ANNz-stack, the Highest Weight Element and the spectroscopic data as shown in Fig. \ref{weighted_comparison_distri_annz_hwe}.  
\subsubsection{Cluster Mass Measurement}
\label{subsubsec:cluster_mass}
Galaxy Clusters are one of the primary tools to probe the $\Lambda$-CDM model \citep[for a review, see e.g.][]{Allen2011}. Cluster masses can be determined by measuring the tangential alignment of gravitationally lensed galaxies\footnote{For a introduction into gravitational lensing we refer to \citet{Bartelmann2001}.} located behind the clusters. The accuracy of these weak lensing mass estimates suffers from uncertainties in the photometric redshift of the lensed sources. In combination with other effects such as cluster mass profile variances, they can introduce systematics at the $5\%$ to $10\%$ level \citep[see e.g.][]{Applegate2014}. In the following, we will only consider uncertainties due to errors in photometric redshift estimates \citep{Seitz1997, Mandelbaum2008, Dawson2012, GruenBrim2013, GruenWISCy2014, Applegate2014}. The excess surface density inside radius $R$ 
\begin{equation}
	\left\langle \Sigma(r) \right\rangle_{r < R} - \overline{\Sigma}(R) = \Sigma_{\rm crit} \, \gamma_{\rm tan}(R)
\end{equation}
is proportional to the critical surface density 
\begin{equation}
\left\langle \Sigma_{\rm cr} \right\rangle \propto \int_{z_{\rm Lens}}^{\infty} dz \, p(z) \left( \frac{D_{d}(z_{\rm Lens}) D_{ds}(z_{\rm Lens}, z)}{D_{s}(z)} \right)
\end{equation}
of the lens at redshift $z_{\rm Lens}$. Here $D_{\rm d}$, $D_{\rm s}$ and $D_{\rm ds}$ denote the angular diameter distance to the lens, the source and between the lens and the source respectively. Uncertainties in the sample PDF of background sources $p(z)$ will propagate into systematic errors in the determination of the critical surface density. This introduces systematic errors in the excess surface density and therefore in the cluster mass estimate.

We quantify the systematic bias of the critical surface density as
\begin{equation}
	{\rm Bias}_{\rm \left\langle \Sigma_{\rm cr} \right\rangle}  = \left( \frac{\left\langle \Sigma_{\rm cr} \right\rangle_{\rm photo} - \left\langle \Sigma_{\rm cr} \right\rangle_{\rm spec}}{\left\langle \Sigma_{\rm cr} \right\rangle_{\rm spec}} \right) \, ,
	\label{eq:def_bias_crit}
\end{equation}
where $\left\langle \Sigma_{\rm cr} \right\rangle_{\rm photo}$ is estimated from the photometry of the objects (e.g., using machine learning) and $\left\langle \Sigma_{\rm cr} \right\rangle_{\rm spec}$ from the spectroscopic redshifts.
\begin{figure}
   \centering
 \includegraphics[scale=0.4]{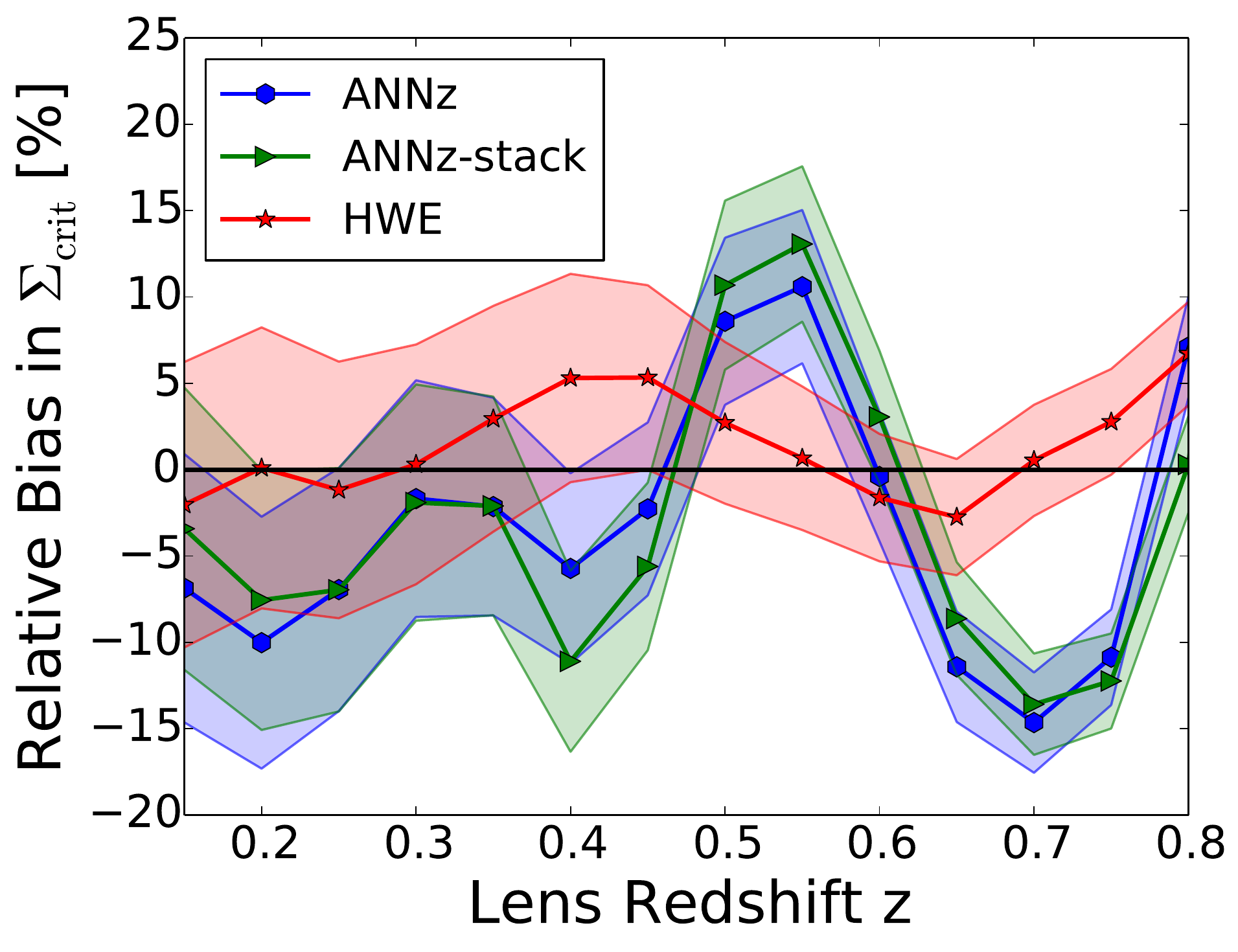}
   \caption{ \label{comparison_sigcrit} Relative bias in the mean critical surface density (Eq. \ref{eq:def_bias_crit}) for different lens redshifts obtained using different estimates for the sample PDF. The filled area shows the $1 \sigma$ error interval.} 
\end{figure}
We estimate the error $\sigma$ on this bias with respect to our test set containing $N$ objects as

\begin{equation}
 \sigma^2 = \left(\frac{\sigma_{\rm photo}(\Sigma_{\rm cr})}{\sqrt{N} \left\langle \Sigma_{\rm cr}\right\rangle_{\rm spec}}\right)^2 \, .
\end{equation}

The mean and standard deviation of the distribution of $\Sigma_{cr}$ are estimated using the probability density function estimates obtained from ANNz and the Highest Weight Element and we present the results in Fig. \ref{comparison_sigcrit}.

The Highest Weight Element estimate for the sample PDF reduces the systematic bias in the critical surface density compared with ANNz by a factor of four for medium lens redshifts $z \in [0.45, 0.6]$. The systematic bias in $\left\langle \Sigma_{\rm cr} \right\rangle$ obtained from the HWE is consistent with zero for lens redshifts $z < 0.7$ and, in general, outperforms the results obtained with ANNz. Higher lens redshifts are however unrealistic for current survey depths.
\subsubsection{Cosmic Shear}
\label{subsubsec:cosmic_shear}
\begin{figure}
   \centering
 \includegraphics[scale=0.4]{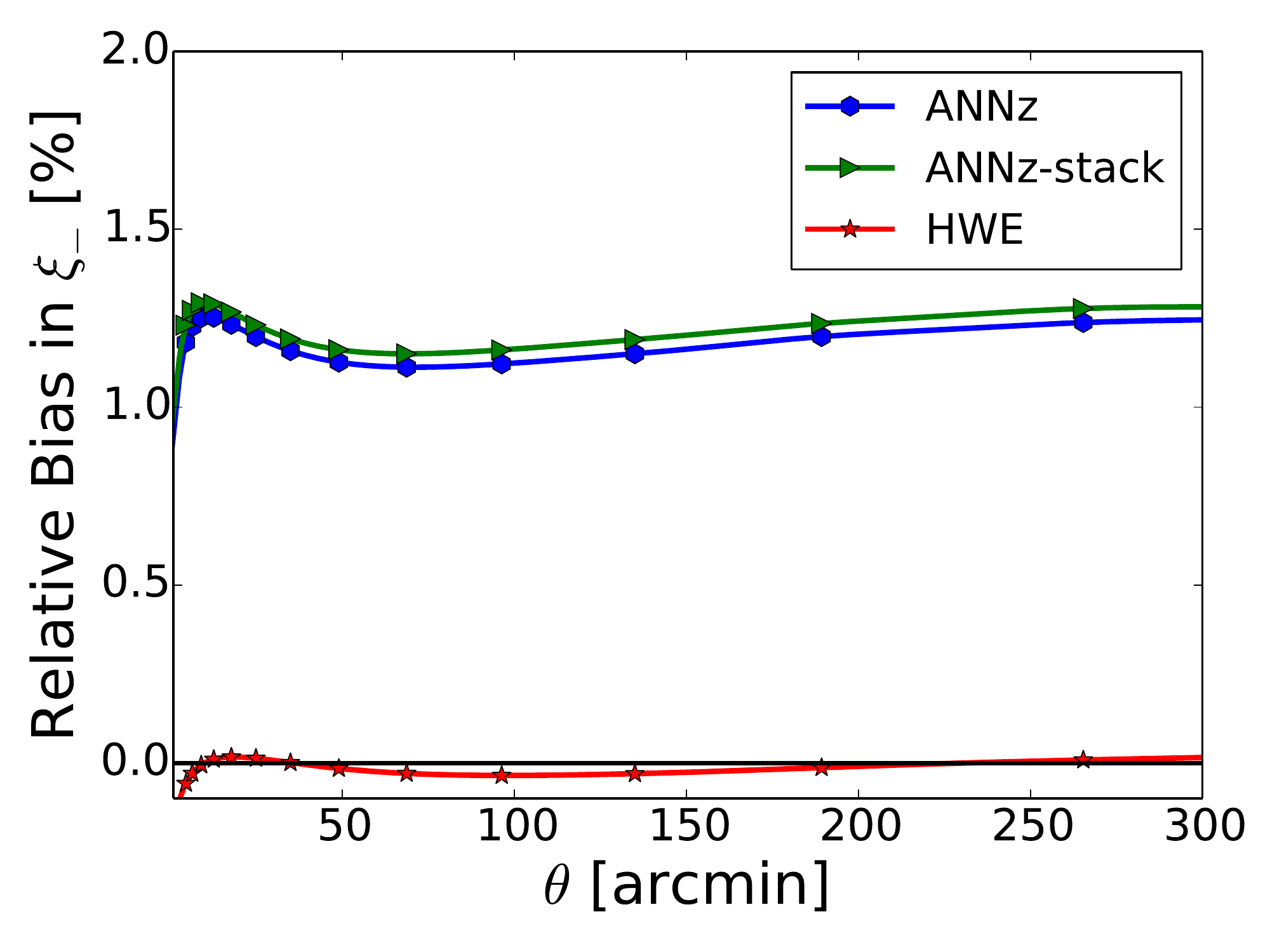}
   \caption{ \label{comparison_lensing_xi_minus} Relative bias in the shear correlation function estimate for $\xi_{-}$ (Eq. \ref{eq:bias_cosmic_shear}) obtained using different estimates for the sample PDF. } 
\end{figure}
\begin{figure}
   \centering
 \includegraphics[scale=0.4]{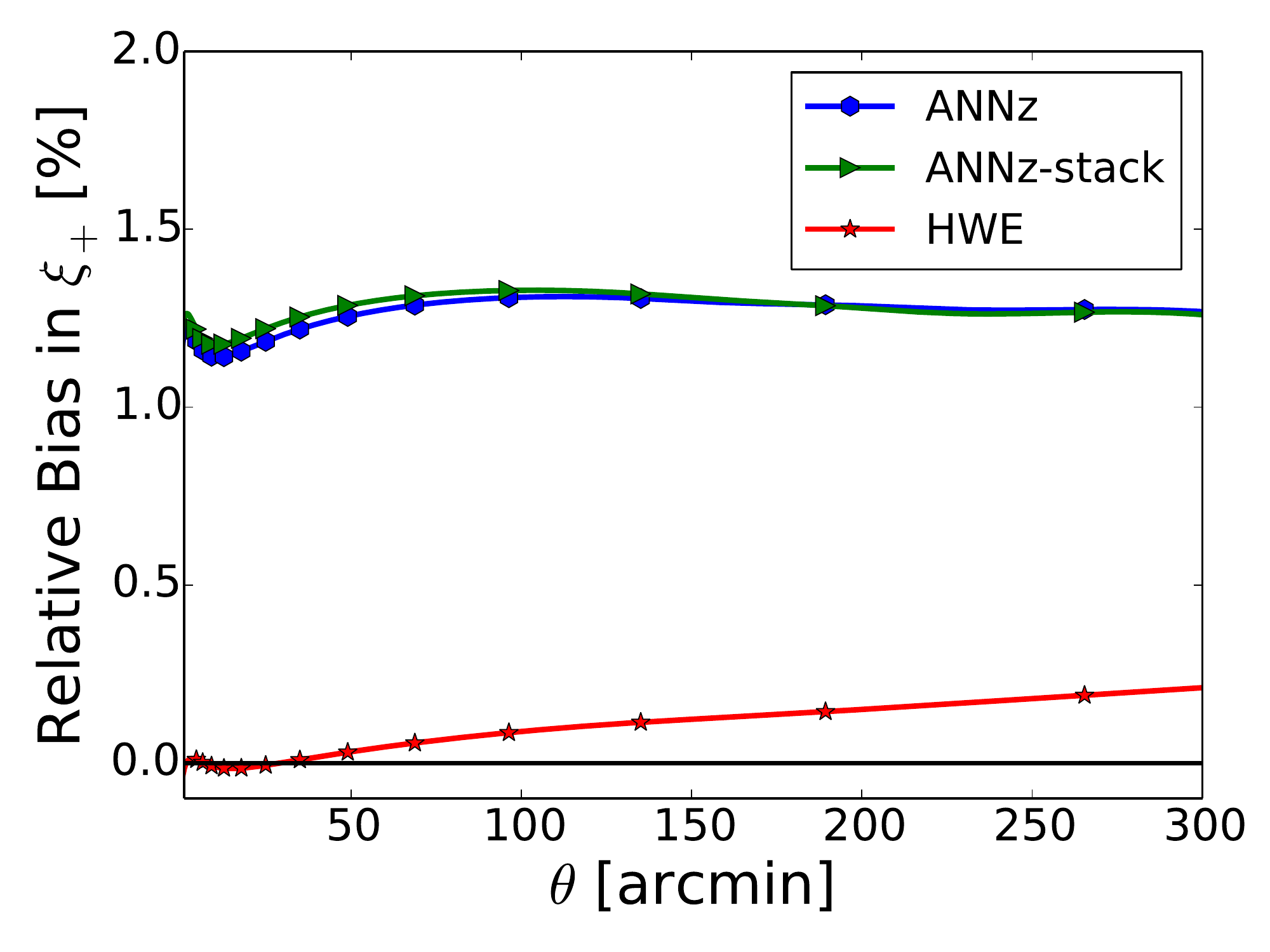}
   \caption{ \label{comparison_lensing_xi_plus} Relative bias in the shear correlation function estimate for $\xi_{+}$ (Eq. \ref{eq:bias_cosmic_shear}) obtained using different estimates for the sample PDF. } 
\end{figure}
Cosmic shear is the weak lensing effect generated by the inhomogenous matter distribution of the universe and has became an important tool to constrain cosmological parameters \citep[see, e.g.,][and references therein]{Kilbinger2013}. Similar to our discussion of the angular correlation function, we derive a power spectrum $P_{\kappa}(\ell)$ of the lensing convergence $\kappa$, which is the source of the lensing potential, defined with respect to the radial co-moving coordinate $x$
\begin{equation}
P_{\kappa}(\ell) = \int_{0}^{\infty} dx \, \left( \frac{q^2(x)}{x^2} \right) P_{\delta}\left(\frac{\ell}{x}, x\right) \, .
\end{equation} 
We calculate the power spectrum $P_{\delta}\left(\frac{\ell}{x}, x\right)$ using the halofit formula from \citet{Smith2003}. The lensing efficiency $q(x)$ quantifies how strongly the objects in an infinitesimal shell of radial comoving coordinates deflect the light coming from background sources. Since the radial comoving coordinates of the objects are related to their redshifts, the lensing efficiency $q(x)$ depends on the sample PDF $p(z)$. From the lensing convergence power spectrum, we can obtain the two shear correlation functions \citep{Kaiser1992} as
\begin{equation}
	\xi_{\pm}(\theta) = \frac{1}{2 \pi} \int_{0}^{\infty} d\ell \, \ell \, P_{\kappa}(\ell) \, J_{0,4}(\ell \theta) \, ,
  \label{eq:xi_plus_minus}
\end{equation}
where the Bessel function $J_0$ ($J_4$) corresponds to the $\xi_{+}$ ($\xi_{-}$) correlation function. In analogy to the previous sections we quantify the bias in the shear correlation functions obtained from photometric data $\xi_{\pm}^{\rm phot}$ by their relative error with respect to the results obtained from the spectroscopic data $\xi_{\pm}^{\rm spec}$, 
\begin{equation}
{\rm Bias}_{\rm \xi_{\pm}} = \left( \frac{\xi_{\pm}^{\rm photo} - \xi_{\pm}^{\rm spec}}{\xi_{\pm}^{\rm spec}} \right) \, .
\label{eq:bias_cosmic_shear}
\end{equation}
The results are presented in Figs. \ref{comparison_lensing_xi_minus} and \ref{comparison_lensing_xi_plus}.
We reduce the bias in the shear correlation function estimates, using the Highest Weight Element estimate instead of the photometric redshift estimates from ANNz, by a factor of 12 for $\xi_{-}$ and a factor of 6 for $\xi_{+}$.

\section{SUMMARY AND CONCLUSIONS}
\label{conclusions}
The next generation photometric surveys will measure the positions on the sky of thousands of millions of galaxies. We must be able to reliably estimate the distance to, or the redshift of, each photometrically identified galaxy before we can use these galaxies in analyses to derive the values of cosmological parameters. Furthermore to maximize the precision and accuracy of any derived parameters, we require a complete understanding of the full shape of the photometric redshift probability density function (hereafter PDF) for both each individual object, and the entire galaxy sample.

In this work we develop and discuss methods drawn from machine learning, to accurately estimate photometric redshift PDFs, which will meet both the future storage demands of large surveys, and the precision demands for cosmological parameter estimation.

As a working example, we apply these algorithms to a sample of galaxies selected from the CFHTLS survey for a set of cosmological analyses. We demonstrate that these methods reduce the biases in all of the analyses examined. We also show that these biases result from the mishandling of the full shape of the photometric redshift PDFs.

This advancement is quantified by comparing several accurate methods to estimate photometric redshift PDFs for individual objects (hereafter individual PDFs). We estimate individual PDFs using a classification scheme that classifies objects into redshift bins and thereby constructs the PDF using the probabilities for bin membership. In contrast to the classification-based PDF estimation methodology commonly used in the astrophysics literature, we incorporate the order of consecutive redshift bins into the classification framework. This produces more accurate individual PDFs. We quantify the performance of the methods by measuring the average log-likelihood of all PDF estimates in a test sample. Our method outperforms other non-ordinal classification and regression schemes, for example classification trees and Neural Networks. Specifically, for high redshift objects, our method reaches performance gains of over 50\% in average log-likelihood when compared with the results obtained using the common Neural Network code ANNz. We construct the individual PDFs using kernel density estimation which inherently requires the selection of a suitable bandwidth to govern the smoothing scale. We propose an efficient method to choose the smoothing scale on an object by object basis. We further discuss a Gaussian mixture model, whose complexity is adaptively selected for each individual object, using a criterion that penalizes model complexity. This method shows solid performance compared with kernel density estimates, while providing a more efficient parametrization of individual PDFs.

Many cosmological analyses require an accurate knowledge of the full shape of the galaxy sample PDF, instead of estimates for the individual PDFs of each galaxy. Sample PDFs are typically obtained by stacking the PDFs of individual galaxies, and so their estimation and storage is required. This reconstruction of the individual PDF typically requires the storage of several hundred floating point numbers. Complex post processing algorithms can reduce this number to 10 - 20 floating point numbers per object at the expense of additional computation time. However in this work, we propose a new single point estimator for each galaxy, called Highest Weight Element (HWE), which can be used to accurately reconstruct the full sample PDF. This leads to a significant reduction in the storage requirements of future photometric surveys. Furthermore, we note that reconstructing the full sample PDF using the point estimator method described in this paper requires orders of magnitude less computation time than using other common redshift codes.

Applications such as shear tomography require the accurate photometric selection of objects in redshift bins. We weight photometrically observed galaxies such that their sample PDF lies within the predefined redshift range. The weights are estimated from the overlap between the individual redshift PDFs and the redshift selection interval. We further use these weights to improve the selection of a sample of galaxies, such that their sample redshift PDF is more accurately confined to be within the predefined redshift bin.

We now return our attention to the specific use case highlighted above using CFHTLS galaxies. In particular we examine the following cosmological analyses: the estimation of cluster masses using weak gravitational lensing, the modelling of galaxy angular correlation functions, and the modelling of cosmic shear correlation functions. In each case we compare the results, and estimate biases, using results obtained with ANNz.

For lensing clusters within the redshift interval $0.45<z<0.6$, we show that our methods reduce the relative bias in the cluster mass reconstruction by up to a factor of 4. Furthermore our methods improve the relative biases in the modelling of the explored large scale structure, and cosmic shear correlation functions by similar values.

In this paper we have shown that the usual point estimate of a photometric redshift is a poor estimator when used to reconstruct the full sample redshift PDF. We note that these point estimates are still used in many recent analyses, and we have shown that their continued use can lead to large biases in cosmological analysis. By using the new HWE point estimator method, highlighted in this paper, we show that the full shape of the sample PDF can be estimated more accurately and that this reduces the biases incurred by mis-estimating the sample PDF. 

The results discussed in this paper have been obtained under the idealized assumption that the data used to train the models is completely representative of the test data. In applications where this is not the case, data augmentation techniques \citep{Hoyle2015} can be used to artificially populate regions of color-magnitude space, that are not fully covered by spectroscopy. These techniques assume a model for the data distribution and can be seen as a form of extrapolation. Weighting methods (\S \ref{subsubsec:catalogue_creation}) are in some cases an alternative to data augmentation. If all relevant attributes are included, these algorithms can be used to determine weights, such that the weighted dataset resembles a reference dataset. 

To aid the common adoption of these tools and techniques we will make the source code of all algorithms publicly available on the homepage of the first author.

\section{ACKNOWLEDGEMENTS}
\label{acknowledgements}
S. Seitz thanks Ofer Lahav for drawing her attention towards machine learning techniques several years ago, and for inspiring discussions on machine learning vs\rc{.} template fitting during mutual visits.
\\
M. M. Rau especially thanks Jolanta Krzyszkowska and Natascha Greisel for useful discussions.
\\
This work was supported by SFB-Transregio 33 `The Dark Universe' by the Deutsche Forschungsgemeinschaft (DFG) and the DFG cluster of excellence `Origin and Structure of the Universe'.

\section*{APPENDIX: TESTS OF WEIGHTING SCHEME}
\label{appendix}
The analyses in \S \ref{subsubsec:cosmic_shear}, \S \ref{subsubsec:cluster_mass} have been carried out by weighting the photospectroscopic dataset such that it resembles a shape catalogue. If only a few objects in the reweighted catalogue are given high weights, the analyses can strongly depend on these objects. We lack spectrocopically observed objects at the faint end of the shape catalogue and therefore employ a magnitude cut to avoid giving large weight to the faint, unrepresentative part of the spectrophotometric catalogue. In analogy to \citet{dessanchez}, we test the robustness of our weighting scheme with respect to the considered applications by excluding the top 5\% of the objects that are given the highest weights.

\begin{figure}
   \centering
 \includegraphics[scale=0.4]{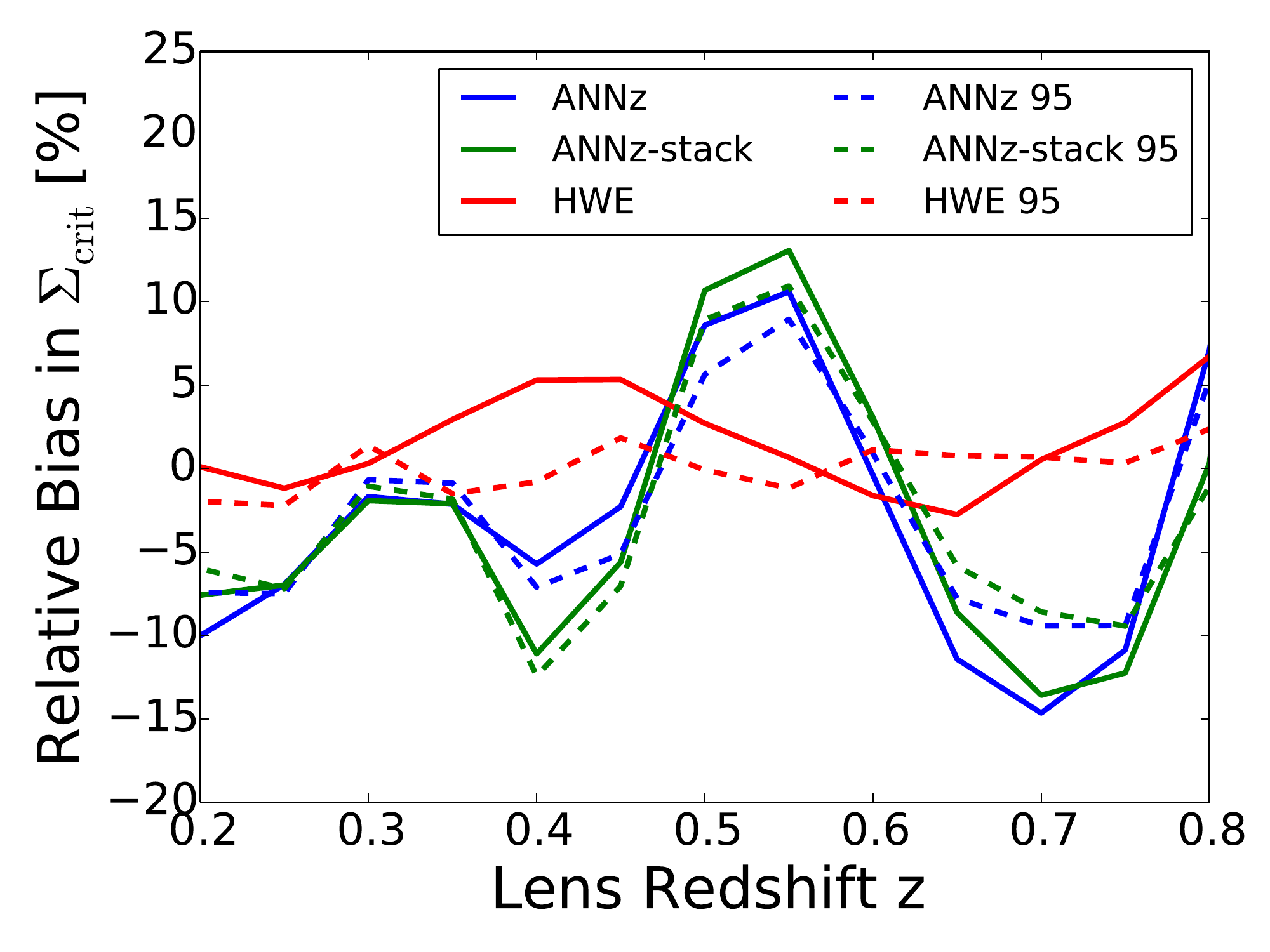}
   \caption{ \label{comp_bias_sigcrit} Relative bias in the mean critical surface density (Eq. \ref{eq:def_bias_crit}) for different lens redshifts obtained using different estimates for the sample PDF. We show the relative biases obtained for the weighted dataset cut at \texttt{MAG\_AUTO} $i' < 23.5$ in solid lines, and the corresponding results with the 5\% highest weighted objects removed in dashed lines.} 
\end{figure}

The bias in the critical surface density is robust against the exclusion of the highest weighted objects for a magnitude cut at \texttt{MAG\_AUTO} $i' < 23.5$ as shown in Fig. \ref{comp_bias_sigcrit}. The results improve for all algorithms if these objects are removed. The conclusions of the analysis, i.e. that the Highest Weight Element (HWE) leads to a lower bias compared with ANNz, remain valid. 

\begin{figure}
   \centering
 \includegraphics[scale=0.4]{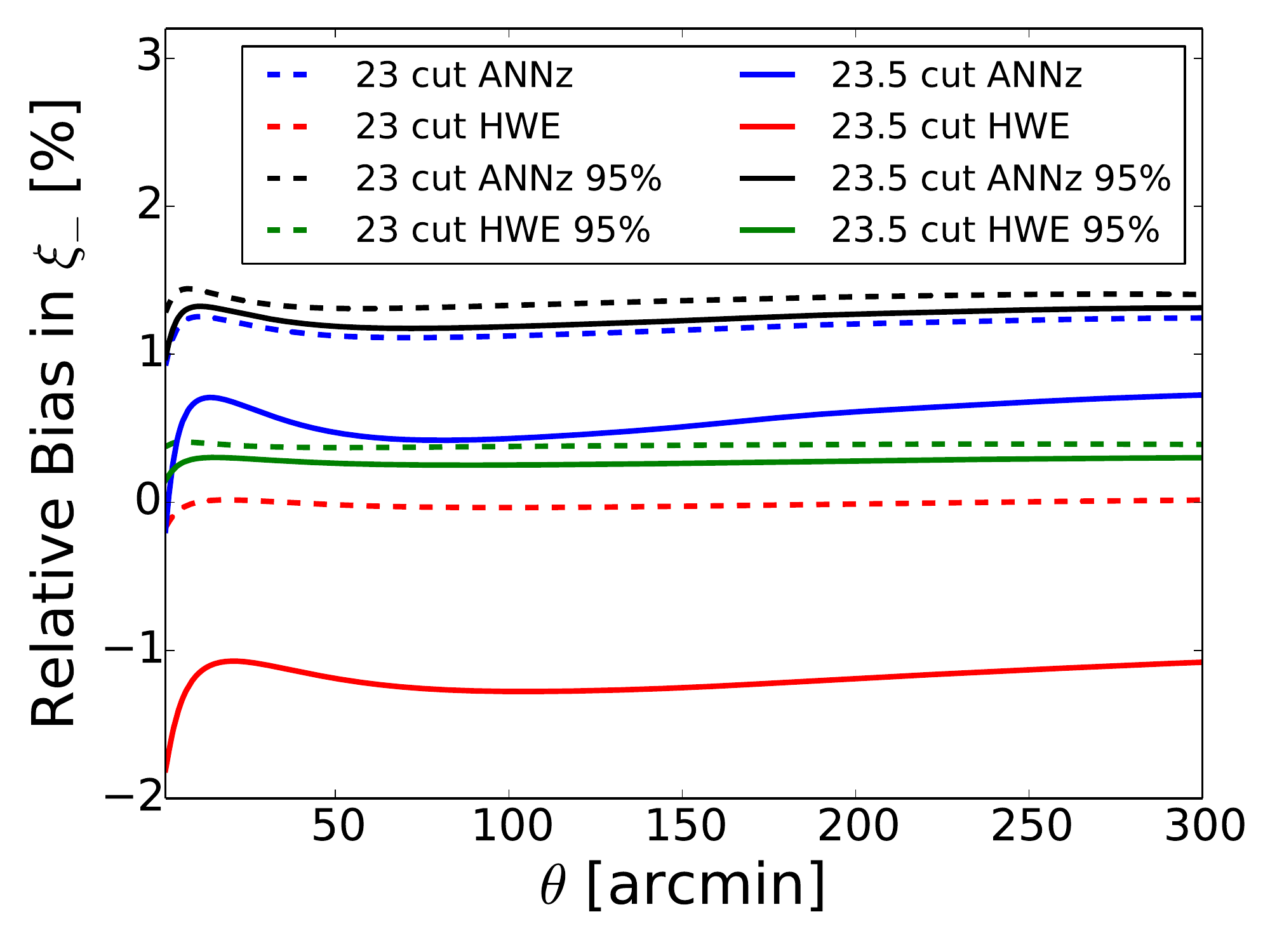}
   \caption{ \label{comp_bias_xi_minus} Relative bias in the shear correlation function estimate for $\xi_{-}$ (Eq. \ref{eq:bias_cosmic_shear}) obtained using different estimates for the sample PDF. We show the relative biases obtained for the weighted dataset cut at \texttt{MAG\_AUTO} $i' < 23.5$ in solid lines and \texttt{MAG\_AUTO} $i' < 23.0$ in dashed lines and the corresponding results with the 5\% highest weighted objects removed.}  
\end{figure}

\begin{figure}
   \centering
 \includegraphics[scale=0.4]{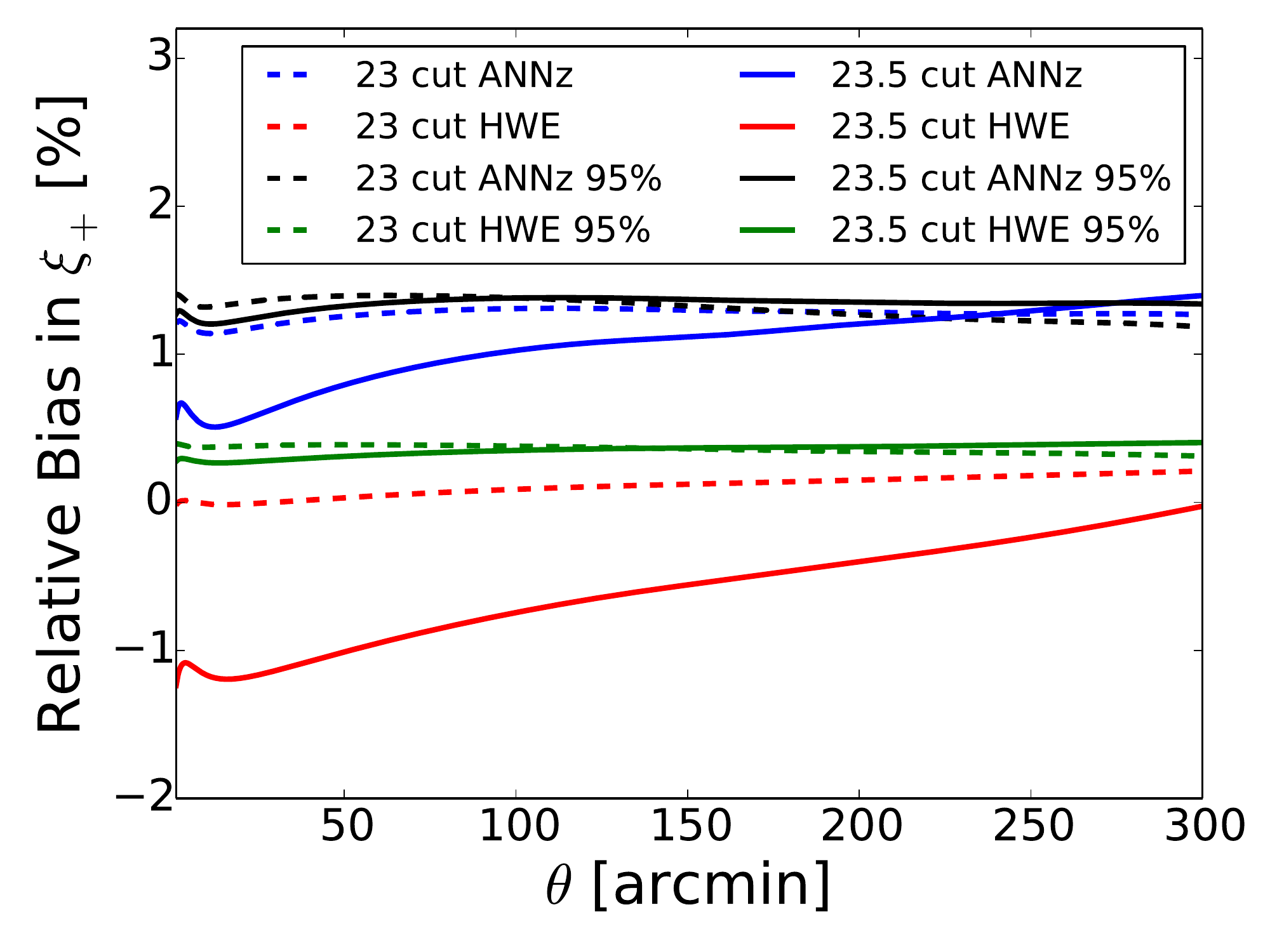}
   \caption{ \label{comp_bias_xi_plus} Relative bias in the shear correlation function estimate for $\xi_{+}$ (Eq. \ref{eq:bias_cosmic_shear}) obtained using different estimates for the sample PDF. We show the relative biases obtained for the weighted dataset cut at \texttt{MAG\_AUTO} $i' < 23.5$ in solid lines and \texttt{MAG\_AUTO} $i' < 23.0$ in dashed lines and the corresponding results with the 5\% highest weighted objects removed.}  
\end{figure}

The analysis of the biases incurred in estimates of the cosmic shear correlation functions requires a more conservative cut at \texttt{MAG\_AUTO} $i' < 23.0$, to be robust against the removal of a small number of highly weighted objects, as can be seen in Fig. \ref{comp_bias_xi_minus} and Fig. \ref{comp_bias_xi_plus}. For a magnitude cut at \texttt{MAG\_AUTO} $i' < 23.5$, ANNz\footnote{The results for ANNz-stack are very similar. Therefore we do not show them here.} gives a better overall result compared with the HWE, while the opposite is true if the 5\% objects with the highest weight are left out. 

\begin{figure}
   \centering
 \includegraphics[scale=0.4]{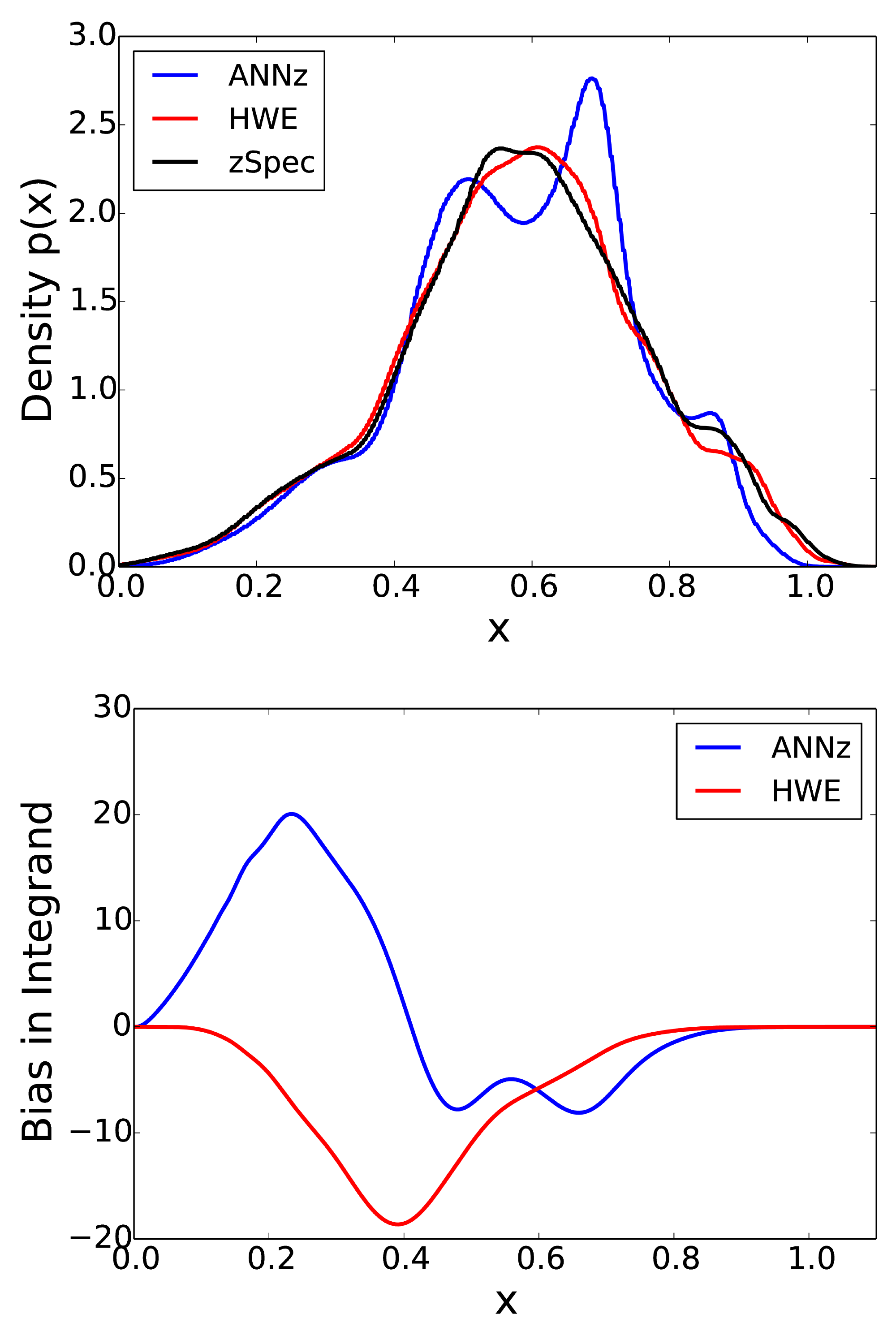}
   \caption{ \label{integrand_explanation} The sample PDFs for a cut at \texttt{MAG\_AUTO} $i' < 23.5$ expressed in radial comoving coordinates $x$ for the spectroscopic, ANNz and HWE reconstructions (top panel). The bias (Eq. \ref{eq:bias_integrand}) obtained with the sample PDFs from ANNz and the HWE for the example $l = 100$ (bottom panel).}     
\end{figure}

Note that this is \emph{not} because the $p(z)$ reconstruction of ANNz is superior at faint magnitudes. 
Instead this can be explained by considering the bias in the integrand in Eq. \ref{eq:xi_plus_minus} with respect to the spectroscopic result given as 
\begin{equation}
	{\rm Bias} = \frac{\ell}{2 \pi} J_{0,4}(l \theta) \left(P_{\kappa}^{\rm phot}(\ell) - P_{\kappa}^{\rm spec}(\ell)\right) \, .
	\label{eq:bias_integrand}
\end{equation}
 As shown in Fig. \ref{integrand_explanation}, ANNz both partly underestimates and overestimates the true spectroscopic integrand at different redshift values such that these two effects compensate each other. Since the lensing efficiency is dominated by the high redshift tail of the stacked PDF, the peculiar shape of the ANNz reconstruction in this range happens to outperform the otherwise superior HWE method. The shape of the high redshift tail strongly depends on a small number of faint objects, which are given a high weight. Accordingly, this artifact is no longer present if the top 5\% of the objects with the highest weights are left out. For a more conservative cut at \texttt{MAG\_AUTO} $i' < 23.0$, the analysis is no longer dominated by a few highly weighted objects at the faint end of our spectrophotometric catalogue, the ANNz analysis does not outperform the HWE, and the interpretation does not depend on the removal of the objects with the highest weights.

\newpage
%BIBLIOGRAPHY
\bibliographystyle{mn2e}
\bibliography{pdf_estim}

\end{document}